\documentclass[referee]{raa}            
\usepackage{graphicx,times}             
\usepackage{natbib}
\usepackage{amssymb,amsmath}
\usepackage{color}
\usepackage{multirow}
\bibpunct{(}{)}{;}{a}{}{,}
\usepackage{threeparttable}

\usepackage[pagebackref=true]{hyperref}

\begin{document}

  \title{A catalog of quasar candidates identified by astrometric and mid-infrared methods in $Gaia$ EDR3
}

   \volnopage{Vol.0 (20xx) No.0, 000--000}      
   \setcounter{page}{1}          

	\author{Qiqi Wu
		\inst{1,2}
		\and
		Shilong Liao\inst{1,2,\ast}
		\and
		Zhaoxiang Qi\inst{1,2}\
		\and
		Hao Luo\inst{1,2}\
		\and
		Zhenghong Tang\inst{1,2}\
		\and
		Zihuang Cao\inst{3}\
	}
	\institute{Shanghai Astronomical Observatory, Chinese Academy of Sciences, 80 Nandan Road, Shanghai 200030, P.R.China
		\and
		University of Chinese Academy of Sciences, No.19(A) Yuquan Road, Shijingshan District, Beijing 100049, P.R.China
		\and
		National Astronomical Observatory, Chinese Academy of Sciences, 20A Datun Road, Chaoyang District, Beijing 100012, P.R.China\\
		\email{shilongliao@shao.ac.cn}
\vs\no
   {\small Received 2022 June 08; accepted 2022 December 09}}

\abstract{Quasars are very important in materializing the reference frame. The excess emission of AGNs (active galactic nuclei) in the mid-infrared band can be used to identify quasar candidates. As extremely distant and point-like objects, quasars also could be further selected by astrometry method. Increasing the number of reliable quasar candidates is necessary in characterizing the properties of $Gaia$ astrometric solution and evaluating the reliability of $Gaia$'s own quasars classification. We identify quasars by using appropriate AllWISE [W1-W2] color and different combinations of astrometric criteria. Together with the contamination and completeness, the magnitude,  astrometric properties,  density distribution,   and the morphological indexes of these selected quasars are evaluated. We obtain a quasar candidate catalog of 1,503,373 sources, which contains 1,186,690 candidates (78.9\%) in common with the $Gaia$ EDR3\_AGN catalog and 316,683 newly identified quasar candidates. The completeness of this catalog is around 80\% compared to LQAC5, and the purity of the overall catalog is about 90\%. We also found that the purity of quasar candidates selected by this method will decrease in the crowded sky area and the region with less WISE observations.
}

   \authorrunning{Qiqi Wu \& Shilong Liao }            
   \titlerunning{A quasar catalog of $Gaia$ EDR3 }  

   \maketitle

%
%
\section{Introduction}
\label{sec:introduction}

Quasars,  known as one type of the active galactic nuclei (AGNs), are extremely distant and point-like objects.  Therefore, quasars are very important in materializing the reference frame.  The Third International Celestial Reference Frame (ICRF3),  which is the realization of the International Celestial Reference System (ICRS) at radio wavelengths,  contains 4588 radio sources obtained with the Very Long Baseline Interferometry (VLBI) \citep{charlot2020third}. The European Space Agency's $Gaia$ mission \citep{prusti2016gaia},  which aims to provide more than one billion accurate determination of proper motions and parallaxes of stars, has already provided more than one million quasar candidates in the visible part of the electromagnetic spectrum.  With a comparable accuracy with VLBI, $Gaia$ is dedicated to establishing a new kinematically non-rotating reference frame in the visible wavelengths with its own astrometric measurements of quasars,  named the $Gaia$ Celestial Reference Frame ($Gaia$-CRF) \citep{mignard2018gaia,refId0}.  Furthermore,  quasars are generally considered as almost zero parallax and proper motion objects, therefore they are vital objects to characterize the astrometric properties, such as the $Gaia$ astrometric solution \citep{liao2021probing,liao2021probing1}. With these concepts in mind,  it is crucial to maximizing the number of quasars in optical wavelengths.

There have been lots of efforts taken to enlarge the number of quasars. Since the first quasar was identified  \citep{1963Natur.197.1040S}, over the past decades,  surveys such as the Large Bright Quasar Survey  \citep{hewett1995large},  the Hamburg Quasar Survey \citep{hagen1995hamburg}, the INT Wide Angle Survey \citep{sharp2001first}, the 2DF Quasar Redshift Survey (2QZ) \citep{croom20042df}  and the quasars from Solan Digital Sky Survey (SDSS) \citep{paris2018sloan,lyke2020sloan} contributed the majority of the quasars identified in the optical wavelengths. Together with the new data released from Large Sky Area Multi-Object Fiber Spectroscopic Telescope (LAMOST) \citep{zhao2012lamost, cui2012large, ai2016large, dong2018large, yao2019large}, the number of quasars discovered in recent years increased rapidly.  These quasars have been compiled into various of catalogs such as Veron-Cetty $\&$ Veron catalog (V$\&$V) \citep{veron2010catalogue}, the Large Quasar Astrometric Catalog (LQAC) \citep{souchay2009construction,souchay2012second,souchay2015third,souchay2017,souchay2019lqac},  the Known Quasars Catalog  \citep{liao2019compilation} and the Million Quasars (Milliquas) catalog \citep{flesch2015half,flesch2017vizier,flesch2021vizier}.   

In spite of the large number of quasars confirmed by their spectra, the number of quasars is far from enough for the establishment of a high-precision celestial reference frame.  Based on the excess emission of the AGN in the mid-infrared band  \citep{lacy2004obscured,richards2006spectral},   the mid-infrared color criteria selections have been proven to be very effective  \citep{mateos2012using,stern2012mid,secrest2015identification,assef2018wise}. With mid-infrared data release from the Wide-field Infrared Survey Explorer (WISE, \citealt{wright2010wide}),  \citet{secrest2015identification} selected about 1.4 million AGN candidates (MirAGN), which contribute the majority of the quasars used to define the $Gaia$-Celestial Reference Frame in $Gaia$ Data Release 2 (DR2)  \citep{lindegren2018gaia,mignard2018gaia}.  Using only photometric and astrometric data, \citet{bailer2019quasar} constructed a supervised classifier based on Gaussian Mixture Models to probabilistically classify extragalactic objects in $Gaia$ DR2, in which 690,000 quasars and 110,000 galaxies candidates are identified.

$Gaia$ Early Data Release 3 (EDR3)  \citep{lindegren2021gaia} provides provisional astrometric and photometric data for more than 1.8 billion sources based on the observations made by $Gaia$.  To enlarge the number of quasars, together with the MirAGN catalog,  17 external quasar or AGN catalogs were cross-matched with $Gaia$ EDR3. These  catalogs include both spectroscopically confirmed quasars and quasar candidates,  such as quasars from 2QZ  \citep{croom20042df}, Roma-BZCAT release 5  \citep{massaro20155th}, R90  \citep{assef2018wise},  the WISE color-selected AGN catalog  \citep{secrest2015identification} and SDSS DR14Q  \citep{paris2018sloan}. Machine learning methods also identified lots of quasar candidates, such as  $Gaia$-unwise  \citep{shu2019catalogues}.  These quasar-like objects, 1,614,173 objects in total,  are available in the $Gaia$ Archive as the table $agn\_cross\_id$ (hereafter EDR3\_AGN catalog)  \citep{lindegren2021gaia}. 

See Table \ref{reliability}, among these 17 external quasar catalogs, we have selected six catalogs with a relatively large number of quasars to investigate the composition of the quasar-like objects in EDR3\_AGN catalog. The $Gaia$-unwise catalog contributes most of the quasar-like objects identified in EDR3\_AGN catalog, and considerable parts of these targets overlap with catalogs such as R90,  the  WISE AGN catalog, and SDSS DR14Q catalog. As very distant objects, non-detectable parallax and proper motion are the basic characteristics of quasars. However, the quasar candidates from the color criteria selection only  and the machine learning method may include a large number of false identification objects. For example, for the $Gaia$-unwise catalog, 2,610,583 objects are matched to the $Gaia$ EDR3, however, only 60.1$\%$  of them meet the astrometric criteria to be identified as quasar-like objects in EDR3\_AGN catalog, see Table \ref{reliability}. So there are many stars and galaxies in the $Gaia$-unwise catalog, which is probably also the case in the quasar-like objects in EDR3\_AGN catalog, especially for the six-parameter solution sample \citep{liao2021probing,liao2021probing1}.   Therefore,  among the 1.6 million quasar-like objects identified in EDR3\_AGN catalog, only 429,249 objects (Frame Rotator Sources, FRS hereafter) are selected to compute the $Gaia$-CRF3.   Compared to $Gaia$ DR2, the systematic residuals of astrometry have been greatly improved in EDR3.  Many studies show that the mean proper motion from the confirmed quasars sample is consistent with zero, and no significant systematic residuals are found in global  \citep{fabricius2021gaia,liao2021probing,liao2021probing1}. Therefore, the astrometric solution in EDR3 is reliable enough to use in quasar selection. The non-detectable parallax and proper motion feature can purify quasar candidates selected by the mid-infrared method, which has been proven quite  effective in our previous quasar selection with APOP \citep{qi2015absolute} and AllWISE data  \citep{guo2018identifying}. 

Furthermore, using particular variability and characteristic SEDs (Spectral Energy Distributions) of quasars, $Gaia$ could identify its own quasar list. In $Gaia$ Data Release 3 (DR3), $Gaia$ has performed its own quasars classification with the low-resolution spectral data, G band magnitude and astrometric parameters \citep{collaboration2022gaia}. However, for the quasars with low-resolution spectroscopy, their photometric signatures are not enough to distinguish them from stars.  As indicated by \citet{claeskens2006identification}, at redshifts $z\textless0.5$, $z\sim 2.5$ and $z>3$,  the white dwarfs, F stars and red dwarfs have similar colors with quasars.   Testing the degree of stellar contamination, i.e. , the quality of quasars from the $Gaia$-data-only classifications convincingly will be quite essential.  Combining different approaches with the near-zero proper motion and parallax, it will be more feasible to distinguish quasars from stellar objects  \citep{mignard2012analysis}.  The join quasar candidates selected by $Gaia$'s astrometric data and mid-infrared color data might be the crucial indicator to test the reliability of such quality.

As the first attempt to use $Gaia$ EDR3 astrometric data to select quasar candidates with the combination of mid-infrared data, this paper aims to provide a reliable quasar candidate list with $Gaia$'s own astrometric data and mid-infrared method. These quasar candidates will play an important role in characterizing the properties of the astrometric solution of $Gaia$ (see, eg: \citealt{fabricius2021gaia,liao2021probing,liao2021probing1}), the establishment of the reference frame in optical wavelength   \citep{mignard2018gaia} and the verification of the $Gaia$ quasar catalog identified by its spectroscopy data.  

In section \ref{sec:data used}, we describe the selection process of quasar candidates. The contamination, completeness, morphological indexes and astrometric properties of our catalog are evaluated in section \ref{sec:Characteristics of the quasar candidate catalog}. In section \ref{discussion}, we discuss the quasar candidates in the Galactic plane and the study of anomalous quasars. Finally, in section \ref{sec:Conclusions}, we summarize our results and give our conclusions.

\begin{table*}[!t]
	\centering
	\caption{Reliability and Contribution of $Gaia$ EDR3 AGN.}
	\begin{threeparttable}
	\begin{tabular}{ccc}
		\hline
		Catalog &Quasars identified in EDR3\_AGN catalog/Object matched & Contribution\tnote{1}  \\ \hline
		Roma-BZCAT &95.6\%(3051/3193)& 0.2\% \\ 
		R90 &81.4\%(1012323/1243053)&62.7\% \\ 
		$Gaia$-unwise&60.1\%(1569680/2610583)&97.2\% \\ 
		WISE AGN&89.4\%(585287/654690)&36.3\% \\ 
		2QZ&99.1\%(25134/25375)&1.6\% \\ 
		SDSS DR14Q&84.0\%(308601/367516)&19.1\% \\ \hline
	\end{tabular}
    \begin{tablenotes}   
    	\footnotesize             
    	\item[1] The total percentage of this column will exceed 100\% because there exits common sources between different catalogs.
    \end{tablenotes}
\end{threeparttable}
\label{reliability}
\end{table*}


\section{Data and selection criteria}
\label{sec:data used}

\subsection{Data Used}
WISE (the Wide-field Infrared Survey Explorer mission,\citealt{wright2010wide}) is a satellite with a 40 cm aperture launched by NASA in 2009, it has four bands at 3.4, 4.6, 12 and 22 $\mu$m (hereafter referred to as W1, W2, W3, and W4, respectively), and has angular resolutions of 6.1, 6.4, 6.5 and 12.0 arcseconds in its four bands, respectively. AllWISE catalog   \citep{cutri2013vizier} is built by combining data from the WISE cryogenic and NEOWISE   \citep{mainzer2011preliminary} post-cryogenic survey, which contains the positions, apparent motions, magnitudes and point-spread function (PSF) profile fit information for about 748 million objects.

$Gaia$ EDR3  \citep{lindegren2021gaia}, which was released at the end of 2020, is the early releases of $Gaia$ Data Release 3 (DR3). $Gaia$ EDR3 contains the 5-parameter (positions, parallaxes, and proper motions) astrometric solution for around 585 million sources and the 6-parameter (positions, parallaxes, proper motions and pseudo-colours\footnote{The pseudocolour is the astrometrically estimated effective wavenumber of the photon flux distribution in the astrometric (G) band, measured in $\mu m^{-1}$ (\url{https://gea.esac.esa.int/archive/documentation/GEDR3/Gaia_archive/chap_datamodel/sec_dm_main_tables/ssec_dm_gaia_source.html}).}) astrometric solution for 882 million sources. The magnitude limit is about G $\approx$ 21 mag at the faint end and about G $\approx$ 3 mag at the bright end. In addition, it also provides the 2-parameter (positions) astrometric solution for around 344 million additional sources. For 5-parameter and 6-parameter sources, both position and parallax uncertainties are less than 0.5 mas at G $\leq$ 20 mag, and about 1.0 $\sim$ 1.3 mas at G = 21 mag, while the proper motion uncertainties are almost less than 0.5 mas/yr at G $\leq$ 20 mag, and 1.4 mas/yr at G = 21 mag.

\subsection{Quasar candidate selection criteria}
\begin{figure}[htbp]
	\centering
	\includegraphics[width=0.7\textwidth]{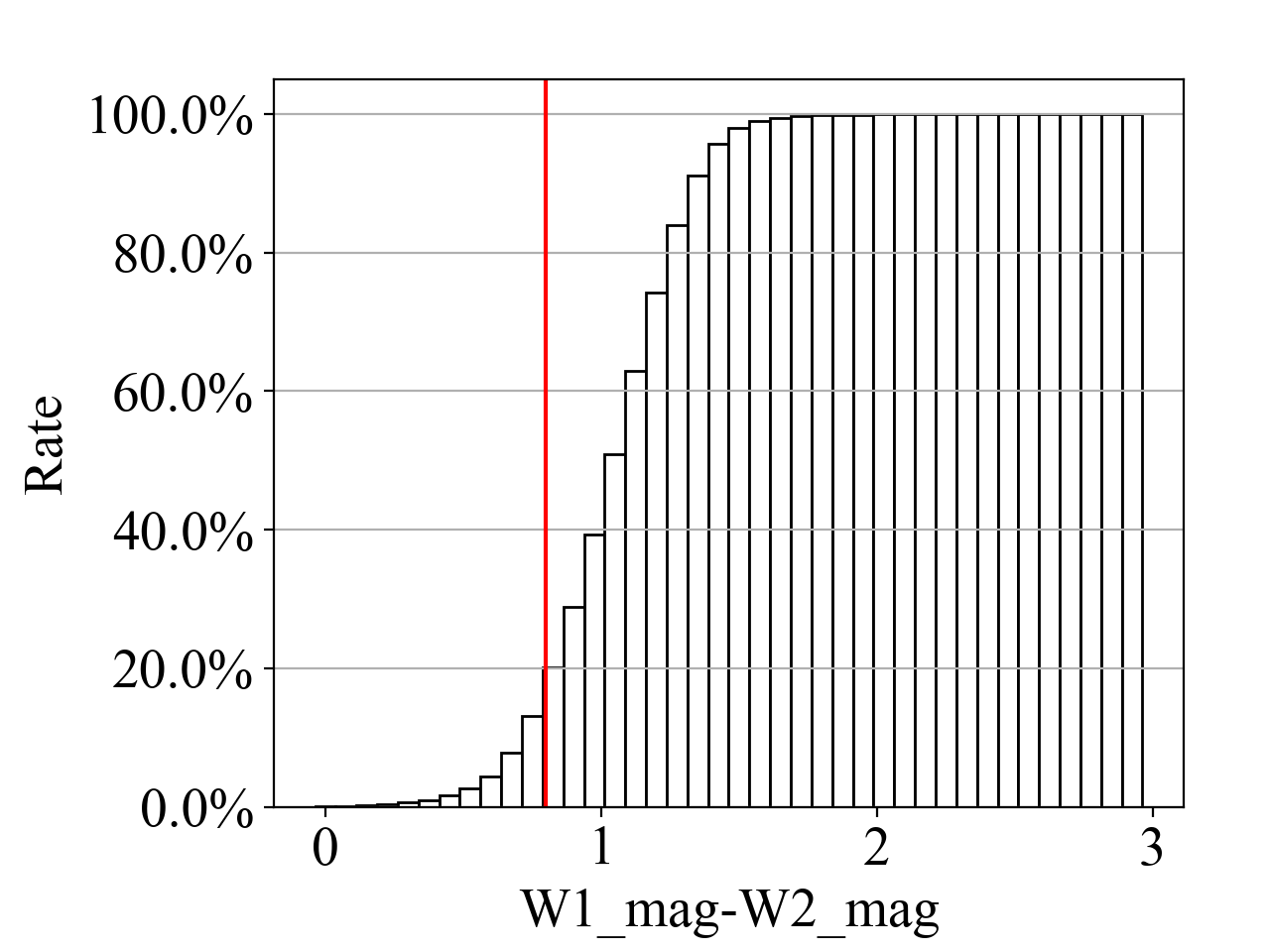}
	\caption{The cumulative histogram for W1-W2 (mag) values of sources in $Gaia$ EDR3 AGN catalog. The red vertical line is W1-W2 = 0.8 mag. }
	\label{agnw1w2}
\end{figure}

With their restricted locus in mid-infrared color space, AGNs can be separated from stars and normal galaxies \citep{lacy2004obscured,stern2005mid,richards2006spectral}.   With the WISE data, \citet{stern2012mid} proposed to use the color information [W1-W2] $\geq$ 0.8 only to select AGNs; \citet{mateos2012using} developed the method of using [W1-W2] and [W2-W3] to define the boundaries of the AGNs located region in mid-infrared color space.  In general, these two methods agree with each other.  We cross-match the $Gaia$ EDR3 AGN catalog with AllWISE catalog and obtain 1,335,906 common sources. Among them, 1,186,690 sources have W1-W2 (mag) values greater than or equal to 0.8, accounting for 88.8\%. The cumulative histogram of W1-W2 (mag) values is shown in Fig. \ref{agnw1w2}, which proves that W1-W2 $\geq$ 0.8 is a reliable criterion for selecting quasar candidates in $Gaia$ EDR3. With the LAMOST DR5 data, our previous study also shows that [W1-W2] $\geq$ 0.8 is a fine balance between low stellar contamination (11.1\%) and high completeness (91.4\%)  \citep{guo2018identifying}. To be consistent with these studies,  we decided to adopt [W1-W2] $\geq$ 0.8 as our mid-infrared color selection criterion. The sources with W1 (or W2) S/N  $\textless$ 5 are rejected to ensure the reliability of our selection results.

For the astrometric criterion,  \citet{lindegren2018gaia} proposed a series of criteria to improve the reliability of the AGN catalog in $Gaia$ DR2.  The matched objects were further selected to have parallaxes and proper motions compatible with zero within five times the respective uncertainty.  Similar criteria are used in the $Gaia$-CRF3 quasars selection \citep{refId0}.

Therefore, based on these studies, we propose the criteria for the selection of quasar candidates in $Gaia$ EDR3 as follows:

\begin{equation}
	\left\{
	\begin{array}{lr}
		(i)\quad  [W1]-[W2] \geq 0.8, & \\
		(ii)\quad  astrometric\_params\_solved = 31\quad  or\quad  95, &  \\
		(iii)\quad  \vert (\varpi + 0.017mas)/\sigma_\varpi \vert \textless 5, &  \\
		(iv)\quad  X^2_{\mu} \equiv  \left[\begin{array}{cc}
		  \mu_{\alpha*} & \mu_{\delta}
		\end{array}\right] Cov(\mu)^{-1} 
			\left[\begin{array}{c}
				 \mu_{\alpha*} \\
				\mu_{\delta} 
			\end{array}\right] \textless 25, &  \\
		(v)\quad  \vert \sin b \vert > 0.1, &  \\
		(vi)\quad  \rho \textless (2arcsec) \times \vert sinb \vert
	\end{array}
	\right.\label{criteria}
\end{equation}
where $b$ is Galactic latitude, $\rho$ is the radius for the positional matching between $Gaia$ EDR3 and AllWISE.  Criterion (ii) selects the objects that have five-parameter or six-parameter solutions; Criterion (iii) takes the  global parallax zero point of EDR3 (-17 $\mu$as)  \citep{lindegren2021gaia1} into consideration; Criterion (iv) adopts the proper motion criteria from \citet{refId0}, where $Cov(\mu)^{-1}$ is the covariance matrix of proper motion; Criterion (v) is designed to avoid dense stars near the Galactic Equator during selection, where it is unreliable to select quasars using mid-infrared method; Criterion (vi) sets the maximum radius for cross-matching, and the combination of (v) and (vi) can effectively improve the accuracy of cross-matching \citep{lindegren2018gaia}.

\begin{figure}[htbp]
	\centering
	\includegraphics[width=0.75\textwidth]{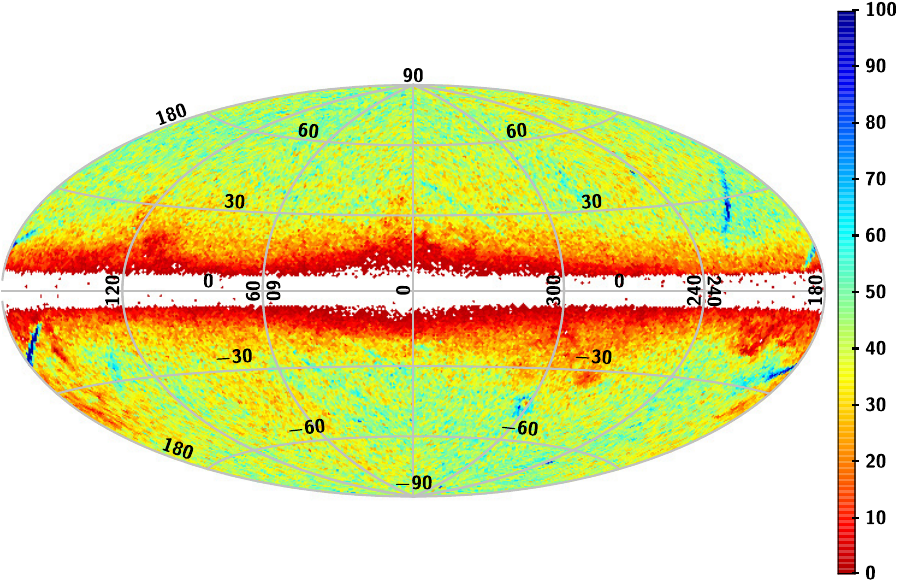}
	\caption{The sky distribution of QCC, using the Hammer Aitoff projection in Galactic coordinate. The cell of this map is approximately 0.84 $deg^2$, and the color shader shows the number of the sources in each cell.}
	\label{denisity}
\end{figure}

With these precepts, we obtain a catalog of 1,503,373 quasar candidates (hereafter as QCC), of which 78.9\% are in the AGN catalog of $Gaia$ EDR3, and other 316,683 sources are newly identified quasar candidates. Table \ref{description} shows the details of our catalog, the astrometric parameters are derived from $Gaia$ EDR3, W1 and W2 magnitude are from AllWISE. Fig. \ref{denisity} shows the density distribution in the sky.

\begin{table*}[!t]
	\centering
	\caption{Description of QCC.  The QCC catalog is available at the CDS.}
	\begin{threeparttable}
	\begin{tabular}{cccccc}
		\hline
		\hline
		Label & Type  & Units & \multicolumn{3}{c}{Detail} \\ \hline
		source\_id & Int   & -     & \multicolumn{3}{c}{Unique $Gaia$ EDR3 source identifier} \\
		WISE\_id & Char  & -     & \multicolumn{3}{c}{Unique AllWISE source identifier} \\
		ra    & Double & degree & \multicolumn{3}{c}{Right Ascension at J2016.0} \\
		ra\_error & Float & mas   & \multicolumn{3}{c}{error of right ascension} \\
		dec   & Double & degree & \multicolumn{3}{c}{Declination at J2016.0} \\
		dec\_error & Float & mas   & \multicolumn{3}{c}{error of right declination} \\
		parallax & Double & mas   & \multicolumn{3}{c}{parallax of the source at J2016.0} \\
		parallax \_error & Float & mas   & \multicolumn{3}{c}{error of parallax} \\
		pm    & Float & mas/year & \multicolumn{3}{c}{total proper motion} \\
		pmra  & Double & mas/year & \multicolumn{3}{c}{proper motion in right ascension direction} \\
		pmra\_error & Float & mas/year & \multicolumn{3}{c}{error of proper motion in right ascension direction} \\
		pmdec & Double & mas/year & \multicolumn{3}{c}{proper motion in declination direction} \\
		pmdec\_error & Float & mas/year & \multicolumn{3}{c}{error of proper motion in declination direction} \\
		pmra\_pmdec\_corr & Float & -     & \multicolumn{3}{c}{Correlation between pmra and pmdec} \\
		ngood\_AL & Short & -     & \multicolumn{3}{c}{number of good observations AL} \\
		G\_mag & Float & mag   & \multicolumn{3}{c}{G-band mean magnitude} \\
		astrometric\_params\_solved & Byte  & -     & \multicolumn{3}{c}{31 for five-parameter solutions, 95 for six-parameter solutions} \\
		W1\_mag-W2mag & Float & mag   & \multicolumn{3}{c}{AllWISE W1 magnitude - W2 magnitude} \\
		Class & Char & -   & \multicolumn{3}{c}{A, B for quasar candidates in QCC-A and QCC-B\tnote{1}}\\ 
		Flag & Char & -   & \multicolumn{3}{c}{ABQ for abnormal astrometric quasar candidates}\\ \hline
	\end{tabular}%
 \begin{tablenotes}   
	\footnotesize             
	\item[1] RQC for the reliable quasar candidates in QCC-B.
\end{tablenotes}
   \end{threeparttable}
	\label{description}%
\end{table*}%

\section{Characteristic of QCC}
\label{sec:Characteristics of the quasar candidate catalog}
In this section,  we investigate the completeness and contamination caused by stars and galaxies of the QCC.   The morphological indexes of the objects in QCC are calculated to further check the reliability. In addition,  we analyze the parallax, magnitude and proper motion distribution of the sources in QCC, and compare them with the AGN catalog in $Gaia$ EDR3.

\subsection{Reliability and Completeness}
\label{reliab}
\begin{figure}[htbp]
	\centering
	\includegraphics[width=0.75\textwidth]{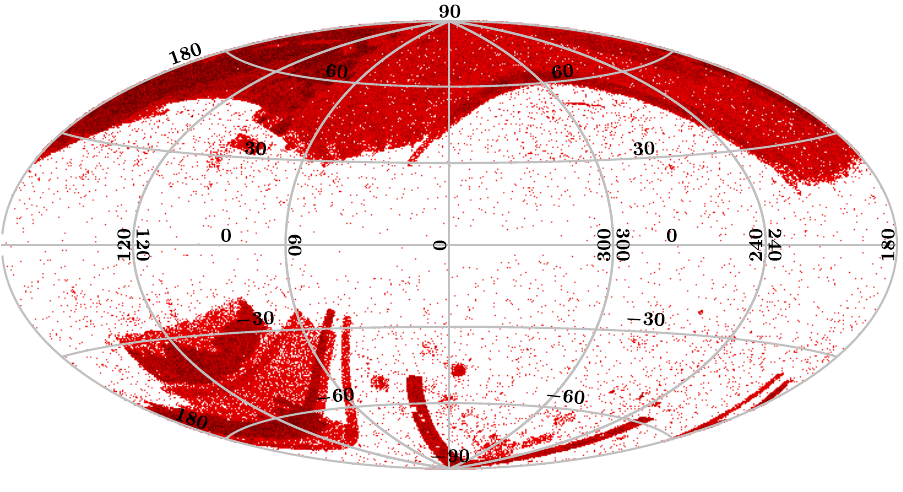}
	\caption{The sky distribution of LQAC5 test catalog, using the Hammer Aitoff projection in Galactic coordinate. }
	\label{lqac5}
\end{figure}

We used the fifth release of the Large Quasar Astrometric Catalog (LQAC-5)   \citep{souchay2019lqac} as a reference to investigate the completeness. To estimate the completeness of QCC, we find 341,987 common sources in LQAC5, $Gaia$ EDR3 and AllWISE. To get a robust sample, we remove  sources with bad AllWISE parameters (S/N in W1 and W2 $\textless$ 5) to obtain a test catalog with 297,527 reliable sources, see Fig. \ref{lqac5}. Among them, 238,807(80.3\%) sources are found in QCC. Table \ref{tab:completeness} shows the completeness of QCC compared with LQAC5. Since we have used strict astrometric criteria for the identifying of quasar candidates, the completeness of QCC is lower than the test catalog, which is about 80\%.

\begin{figure}[h]
	\centering
	\begin{tabular}{c}
		\includegraphics[width=0.25\textwidth]{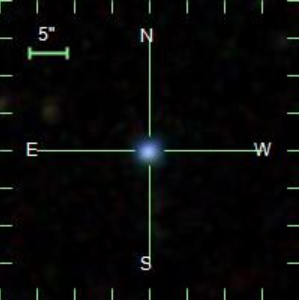}
		\small(A)
	\end{tabular}
	\begin{tabular}{c}
		\includegraphics[width=0.25\textwidth]{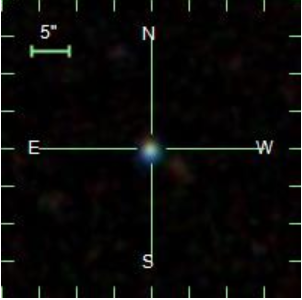}
		\small(B)
	\end{tabular}
	\begin{tabular}{c}
		\includegraphics[width=0.25\textwidth]{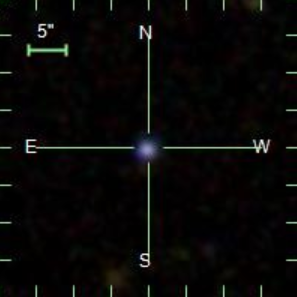}
		\small(C)
	\end{tabular}
	
	\begin{tabular}{c}
		\includegraphics[width=0.25\textwidth]{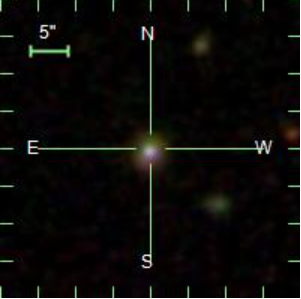}
		\small(D)
	\end{tabular}
	\begin{tabular}{c}
		\includegraphics[width=0.25\textwidth]{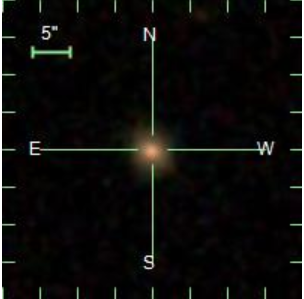}
		\small(E)
	\end{tabular}
	\begin{tabular}{c}
		\includegraphics[width=0.25\textwidth]{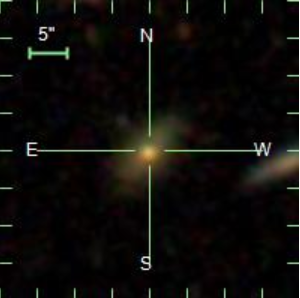}
		\small(F)
	\end{tabular}
	\caption
	{\small Six sources of our catalog matching with SDSS DR16. Panel A, B, C: point-like source, panel D, E, F: extended source.}
	\label{fig:point}%
\end{figure}

To evaluate the contamination caused by stars and galaxies, we randomly select a $10^{\circ} \times 10^{\circ}$ region with a center coordinate of Ra = $255^{\circ}$ and Dec = $35^{\circ}$. In this test region, 4171 common sources are found in QCC and SDSS DR16  \citep{ahumada202016th}, among them, 3230 sources are identified as quasars  in SDSS DR16Q  \citep{lyke2020sloan}. For the remaining 941 sources, we checked the SDSS spectrums of them, the result shows that 1, 9 and 7 of them are classed as star, galaxy and quasar respectively,  with no spectrums from SDSS are available for the rest 924 objects. Therefore,  for all sources that have spectral classifications, quasars  accounted for 99.7\% (3237/3247), while stars and galaxies accounted for 0.3\% (10/3247), which shows that our quasar  candidate catalog identified by astrometry and mid-infrared methods has a very low proportion of contamination. To establish an accurate celestial reference frame, the sources in our final catalog should be point-like sources, Fig. \ref{fig:point} shows the SDSS thumbnails of six candidates in our catalog. Using the same reference region, among the 4171 common sources, 3920 (93.98\%) are found to be point-like objects \footnote{The extended and point-like sources are classified by SDSS morphological data, more details could be found in \url{https://www.sdss.org/dr12/algorithms/classify/}.}, with 251 (6.02\%) identified as extended sources. 

\begin{figure}[htbp]
	\centering
	\includegraphics[width=0.7\textwidth]{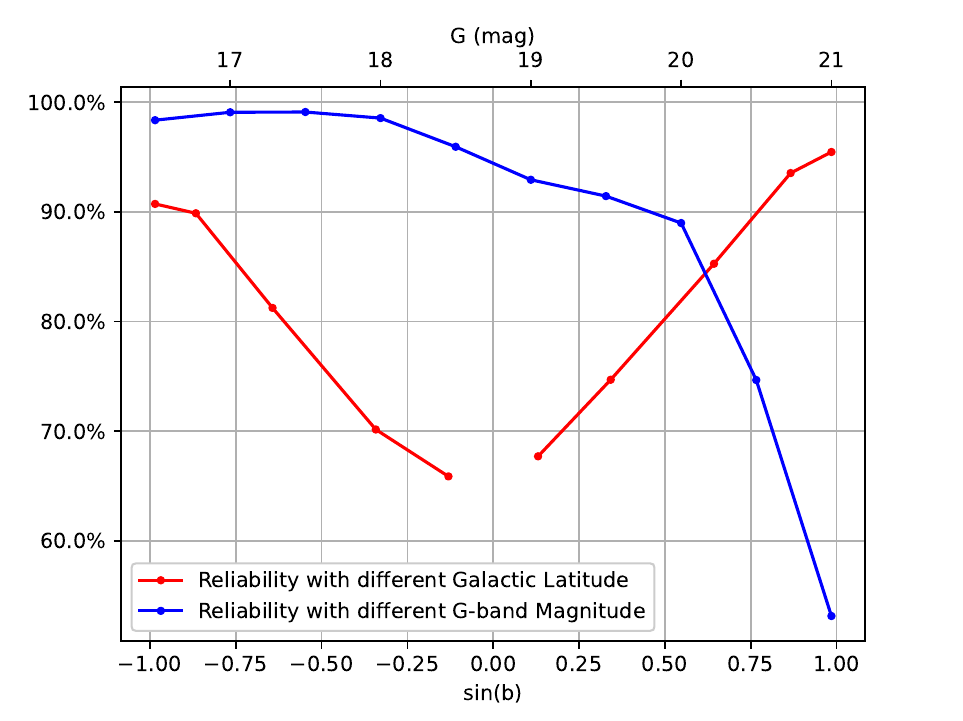}
	\caption{The purity change of QCC. The blue and red line represent the variation of reliability with G magnitude and Galactic latitude, respectively.}
	\label{puremag}
\end{figure}

\begin{figure}[htbp]
	\centering
	\includegraphics[width=0.7\textwidth]{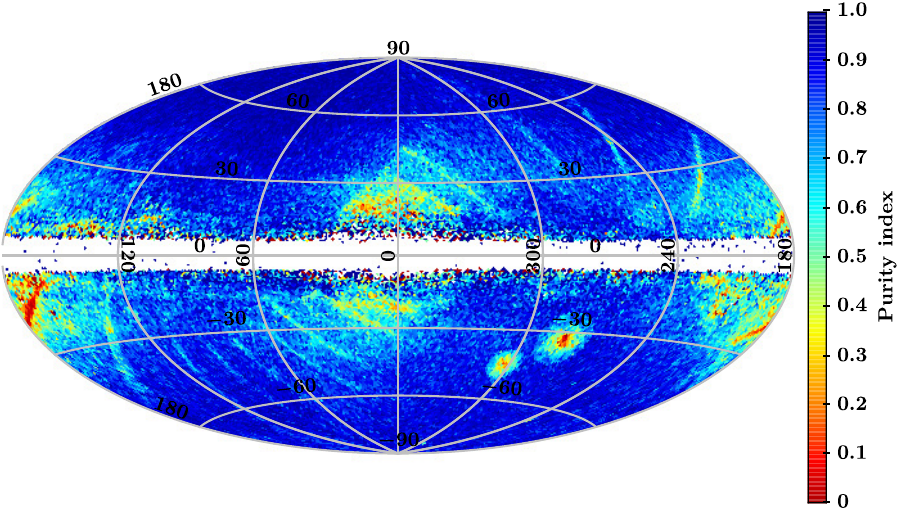}
	\caption{The purity sky distribution of QCC. The map shows the sky density with each cell of approximately 0.84 $deg^2$, using the Hammer Aitoff projection in Galactic coordinate  with zero longitude at the center.}
	\label{puredenisity}
\end{figure}

The purity of quasar candidates in a random sky region illustrates the effectiveness of the method. However,  the purity may vary widely in different sky regions, just as mentioned in \citet{refId0}. Therefore, it is necessary to investigate the purity variation of QCC in the whole sky. To carry out this investigation, we need a high-completeness pure quasar catalog as our reference sample. \citet{collaboration2022gaia} provide a sample of 1.9 million quasar candidates in their Section 8, and the purity of this sample is approximately 95\%. We assume that this sample has 100\% reliability and 100\% completeness, and explore the purity change of QCC by comparing QCC with it. This assumption could indeed be inaccurate, yet convenient to assess how the purity of QCC varies with sky density, Galactic latitude and magnitude. We find that 1,241,033 (82.5\%) quasar candidates of QCC are in this pure sample provided by $Gaia$ DR3. Fig. \ref{puremag} shows the purity change with different Galactic latitude and G magnitude. There is a decrease in the purity of QCC where the sources located near the Galactic Equator and at the fainter magnitude. To explore the correlation between purity change and sky density, we propose the concept of purity index. The purity index indicates the number of common sources between the QCC and the reference catalog divided by the number of sources in QCC in each HEALPix \citep{gorski1999healpix} sky pixel. See Fig. \ref{puredenisity}, QCC has relatively low purity near the Galactic Equator and in the area of the Large and Small Magellanic Clouds (LMC and SMC). Although this does not accurately represent the purity distribution of QCC in the sky, it has important implications for our better understanding of quasar selections by mid-infrared and astrometric method.

\begin{figure}[htbp]
	\centering
	\includegraphics[width=0.7\textwidth]{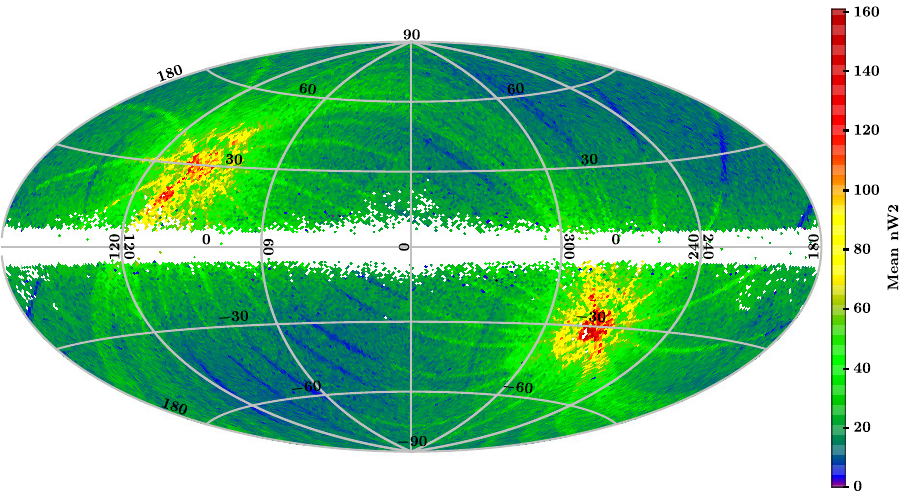}
	\caption{The sky distribution of the integer frame detection count in W2 band for QCC sources. The map shows the sky density with each cell of approximately 0.84 $deg^2$, using the Hammer Aitoff projection in Galactic coordinate with zero longitude at the center.}
	\label{nw}
\end{figure}

The lower purity of QCC near the LMC, SMC and the Galactic Center is to be expected, as these regions are very crowded. Selection of quasar candidates in these regions may require more stringent criteria. In addition to these areas, we found that the purity of QCC in some striped areas and anti-Galactic center areas is also relatively low, see Fig. \ref{puredenisity}. Fewer AllWISE observations may be the main reason for the low purity in these regions. We can find that the area with fewer observations of WISE W2 band in Fig. \ref{nw}\footnote{The detection count represents the number of individual 7.7s exposures on which the source was detected with SNR$\textgreater$3 in the WISE profile-fit measurement.} coincides well with the low-purity area in Fig. \ref{puredenisity}. In summary, when selecting quasar candidates using the method mentioned in this paper, the purity of samples will decrease in crowded sky regions and in the regions with fewer WISE observations.

\begin{table}[htbp]
	\centering
	\caption{Completeness compared with LQAC5. The ``ALL quasars"  represents the number of common sources of each catalog and the LQAC5 test catalog, the ``Quasars found" means how many QCC quasars are found in the ``ALL quasars".}
	\begin{tabular}{ccc}
		\hline
		\hline
		Resource & Completeness & Quasars found/ALL quasars \\ \hline
		ICRF3 & 81.07\% & 2351/2900 \\
		FIRST & 75.27\% & 10462/13899 \\
		2QZ   & 86.65\% & 17181/19827 \\
		SDSS DR12 & 80.20\% & 224440/279845 \\
		GSC2.3 & 80.21\% & 234142/291901 \\
		2MASS & 71.40\% & 19122/26783 \\
		V\&V  & 80.97\% & 75880/93717 \\
		R90   & 91.47\% & 184921/202159 \\
		Roma-BZCAT & 70.73\% & 1839/2600 \\
		$Gaia$ EDR3 FRS & 91.66\% & 93947/102493 \\
		\hline
	\end{tabular}%
	\label{tab:completeness}%
\end{table}%

\subsection{The morphological indexes}
Inspired by LQAC2  \citep{souchay2012second}, we analyzed the morphological indexes of QCC. We apply the photometry function of IRAF to the PSF of each source and compare it to the PSFs of other stars near that source. The optical images used in the calculation process came from SDSS, with a total of 4 bands from blue to infrared, namely g, r, i, z, and the corresponding central wavelengths are 477.0, 623.1, 762.5, and 913.4 nm. The parameters SHARP, SROUND, GROUND determined by IRAF's DAOFIND provide comprehensive morphological data of each source. The morphological indexes of each source are calculated by \citet{andrei2012morphology}:
\begin{equation}
	M_{PC} = \vert P_Q - \overline{P_s} \vert / \sigma_{s}
\end{equation}
where $M_{PC}$ is the morphological index of each source for the parameter P in the colour C, $P_Q$ represents the parameter P of quasar Q, $\overline{P_s}$ is the mean value of parameter P of the stars on the same SDSS field as quasar Q, $\sigma_{s}$ is the standard deviation of these stars' parameters.

\begin{figure}[h]
	\centering
	\begin{tabular}{c}
		\includegraphics[width=0.42\textwidth]{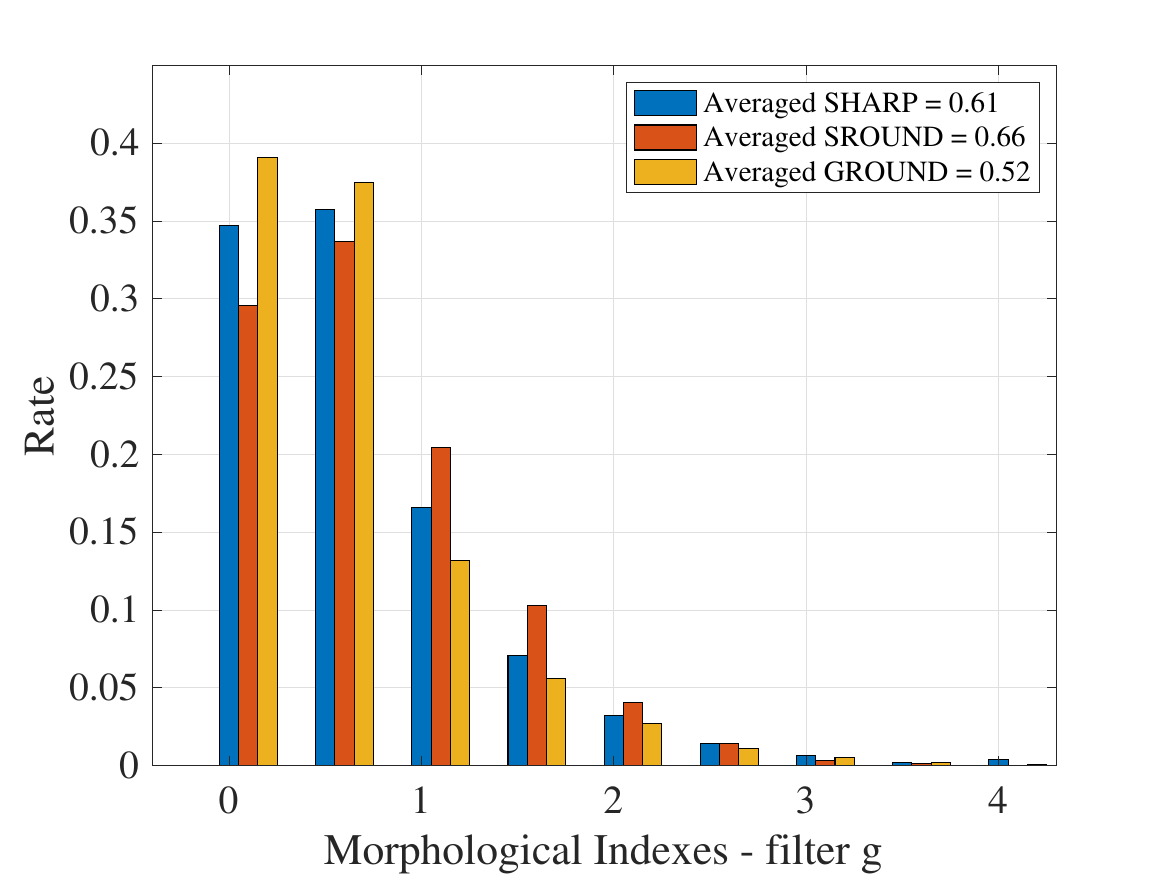}
		\small(A)
	\end{tabular}
	\begin{tabular}{c}
		\includegraphics[width=0.42\textwidth]{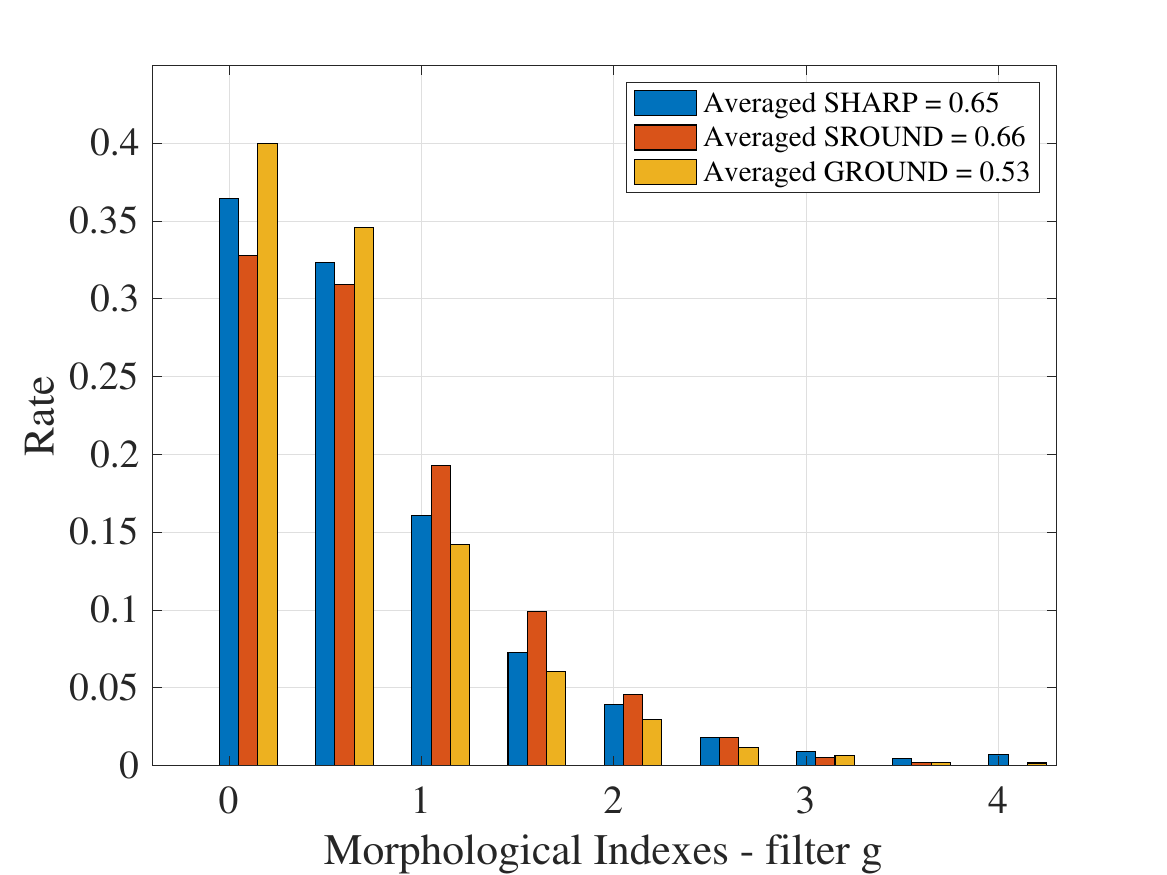}
		\small(B)
	\end{tabular}

   	\begin{tabular}{c}
   	\includegraphics[width=0.42\textwidth]{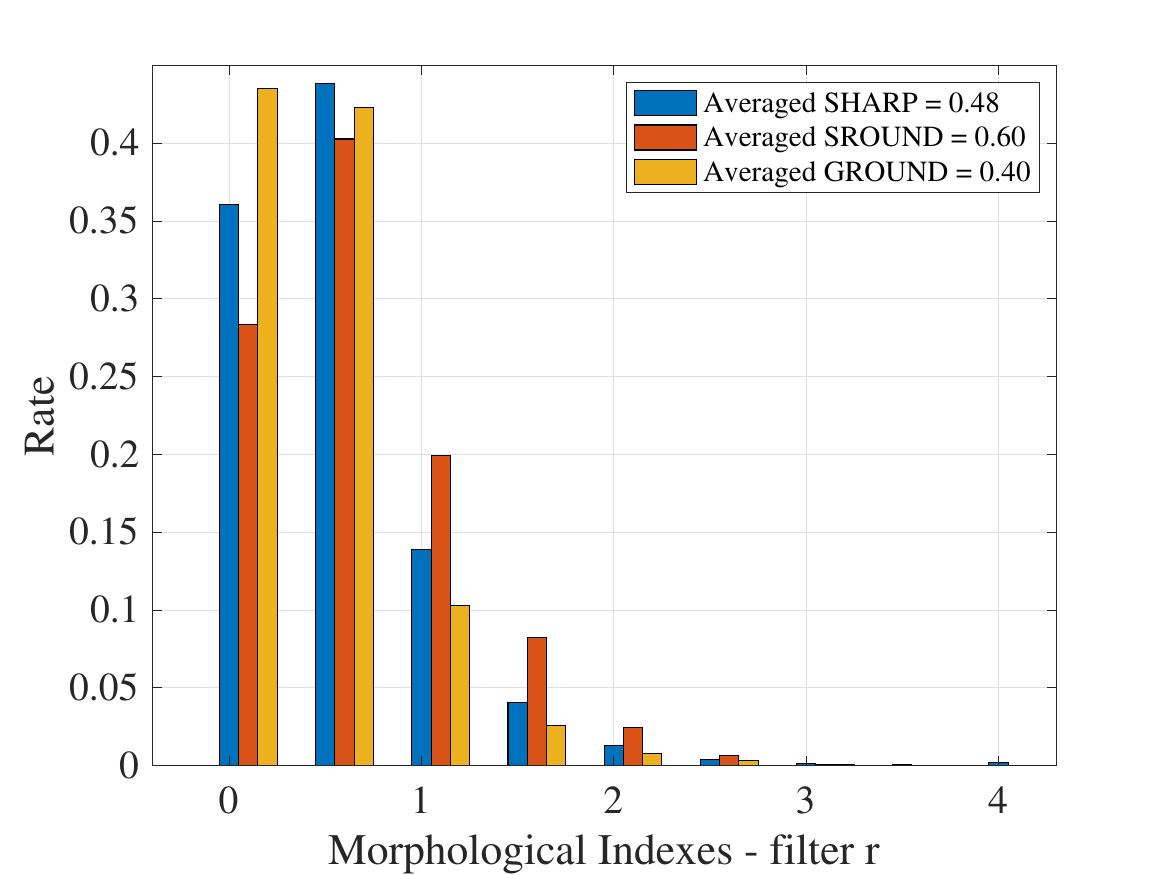}
   	\small(C)
   \end{tabular}
   \begin{tabular}{c}
   	\includegraphics[width=0.42\textwidth]{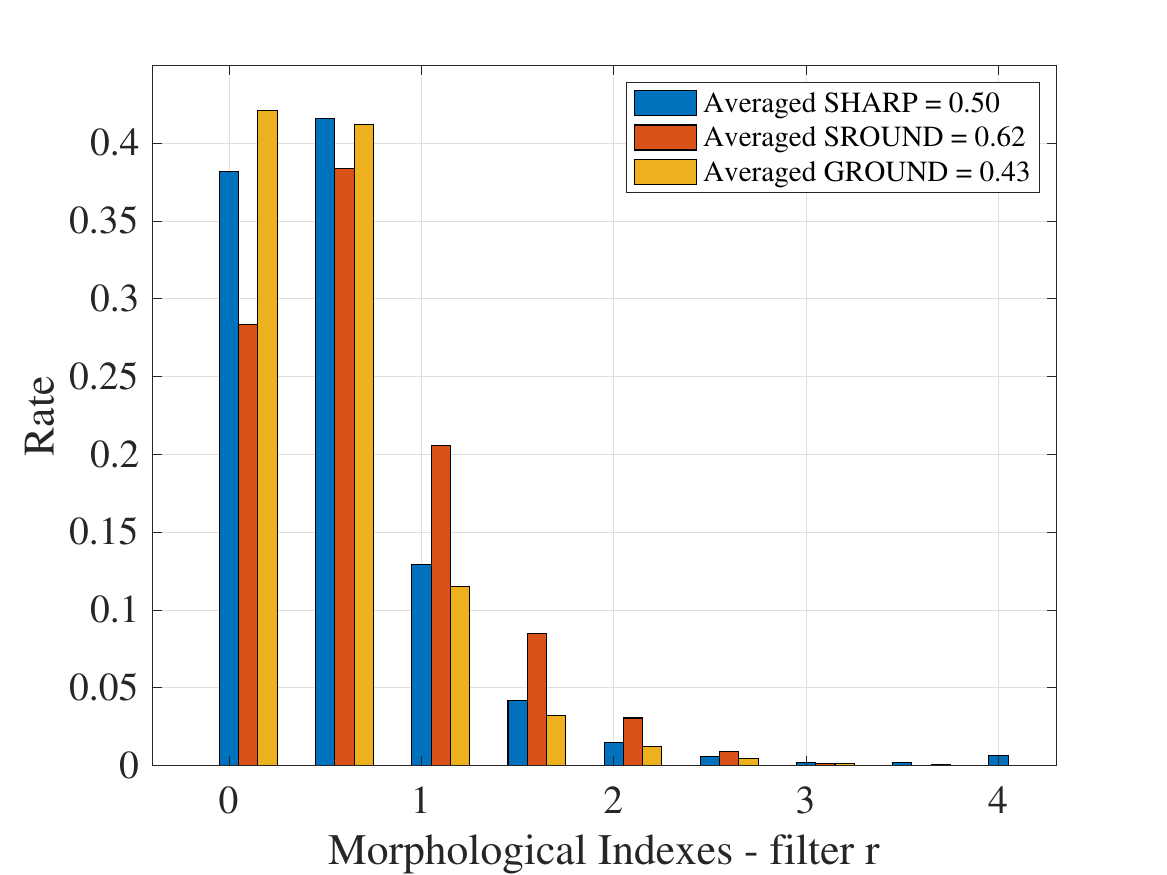}
   	\small(D)
   \end{tabular}

	\begin{tabular}{c}
	\includegraphics[width=0.42\textwidth]{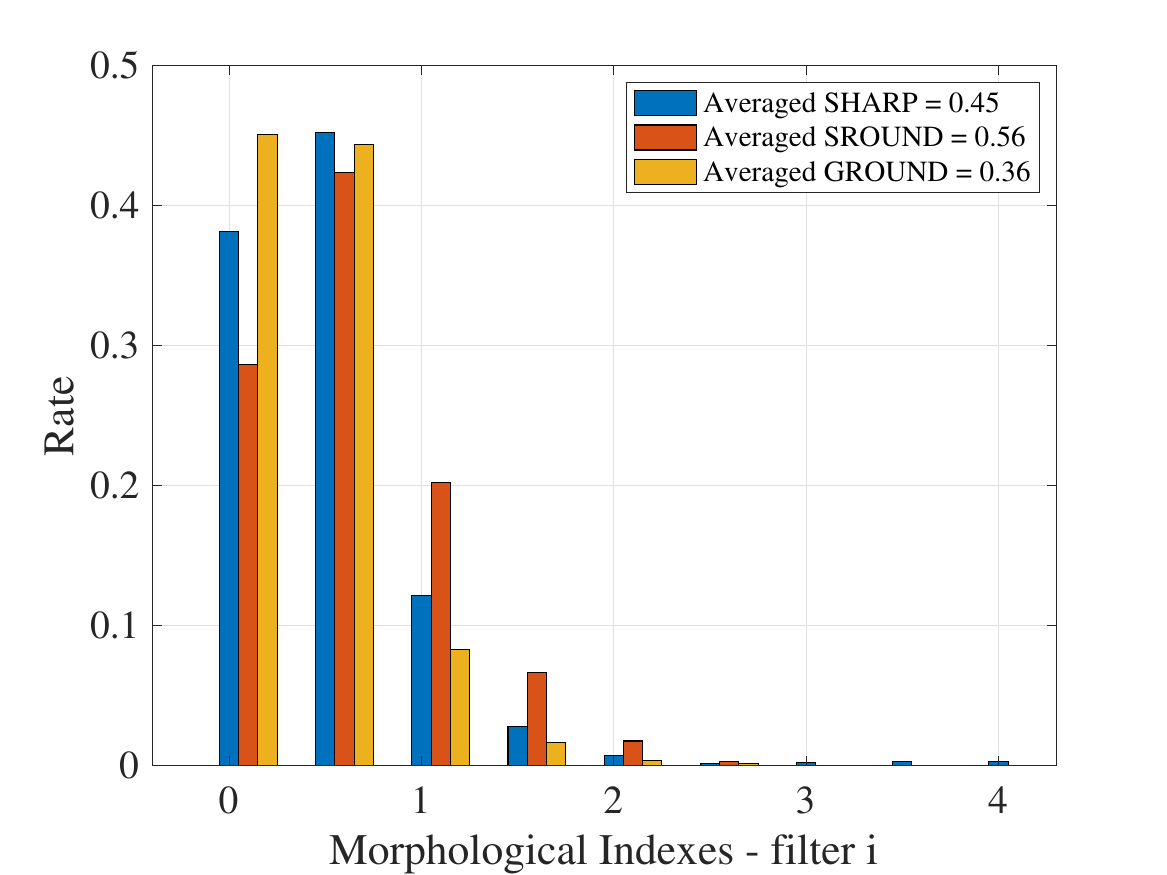}
	\small(E)
\end{tabular}
\begin{tabular}{c}
	\includegraphics[width=0.42\textwidth]{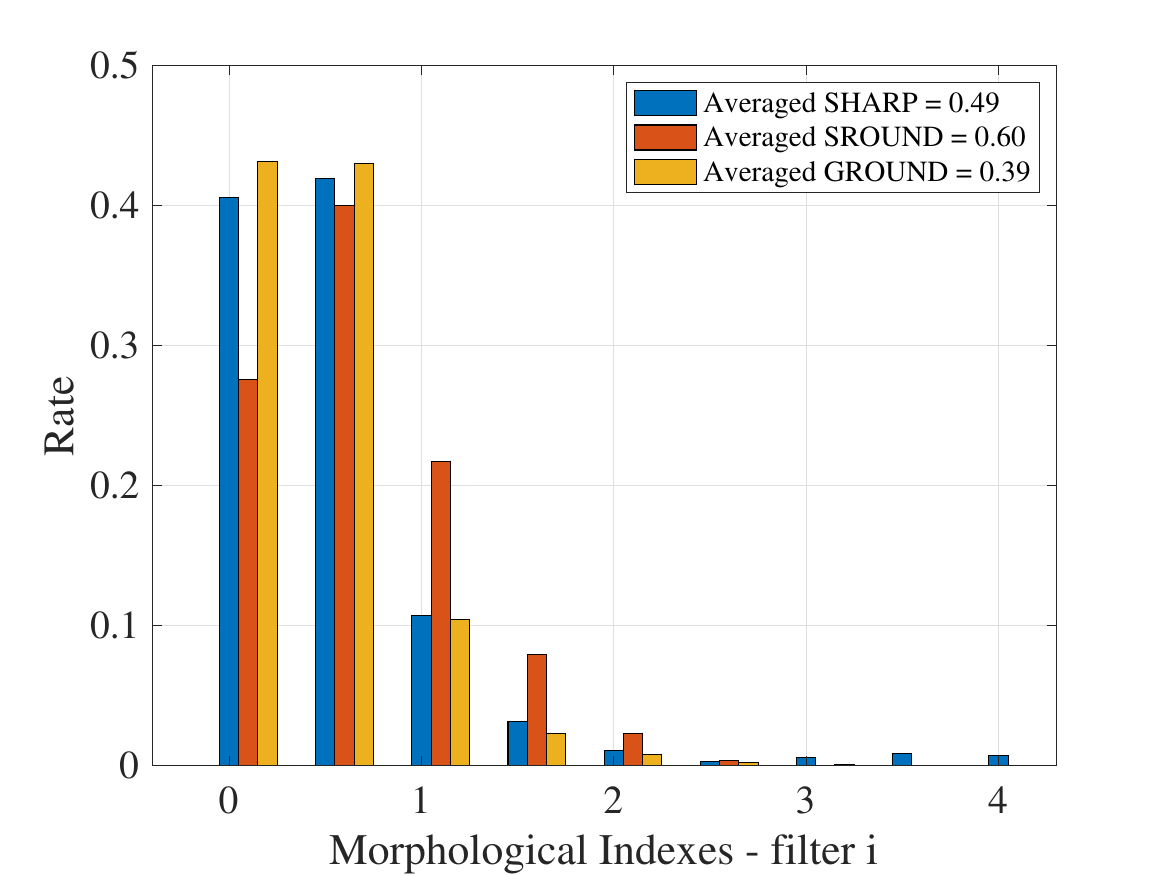}
	\small(F)
\end{tabular}

	\begin{tabular}{c}
	\includegraphics[width=0.42\textwidth]{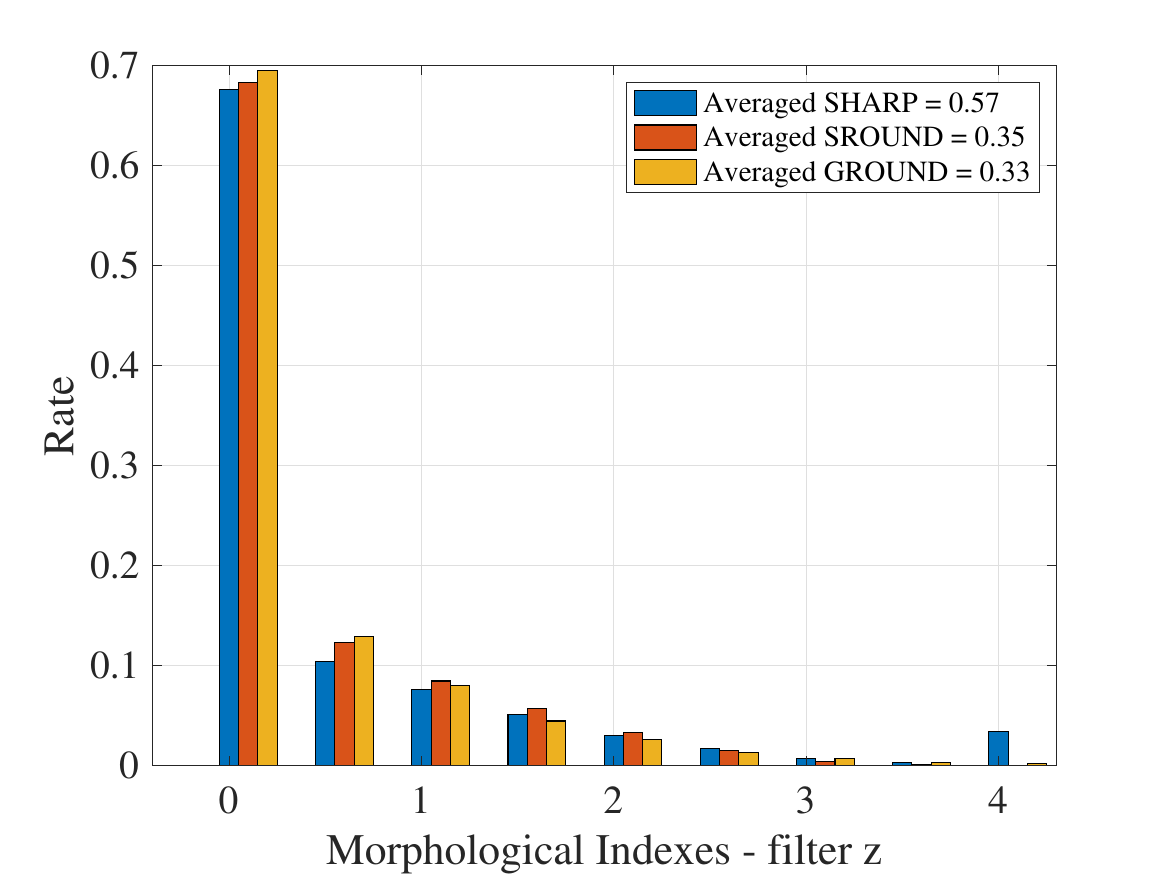}
	\small(G)
\end{tabular}
\begin{tabular}{c}
	\includegraphics[width=0.42\textwidth]{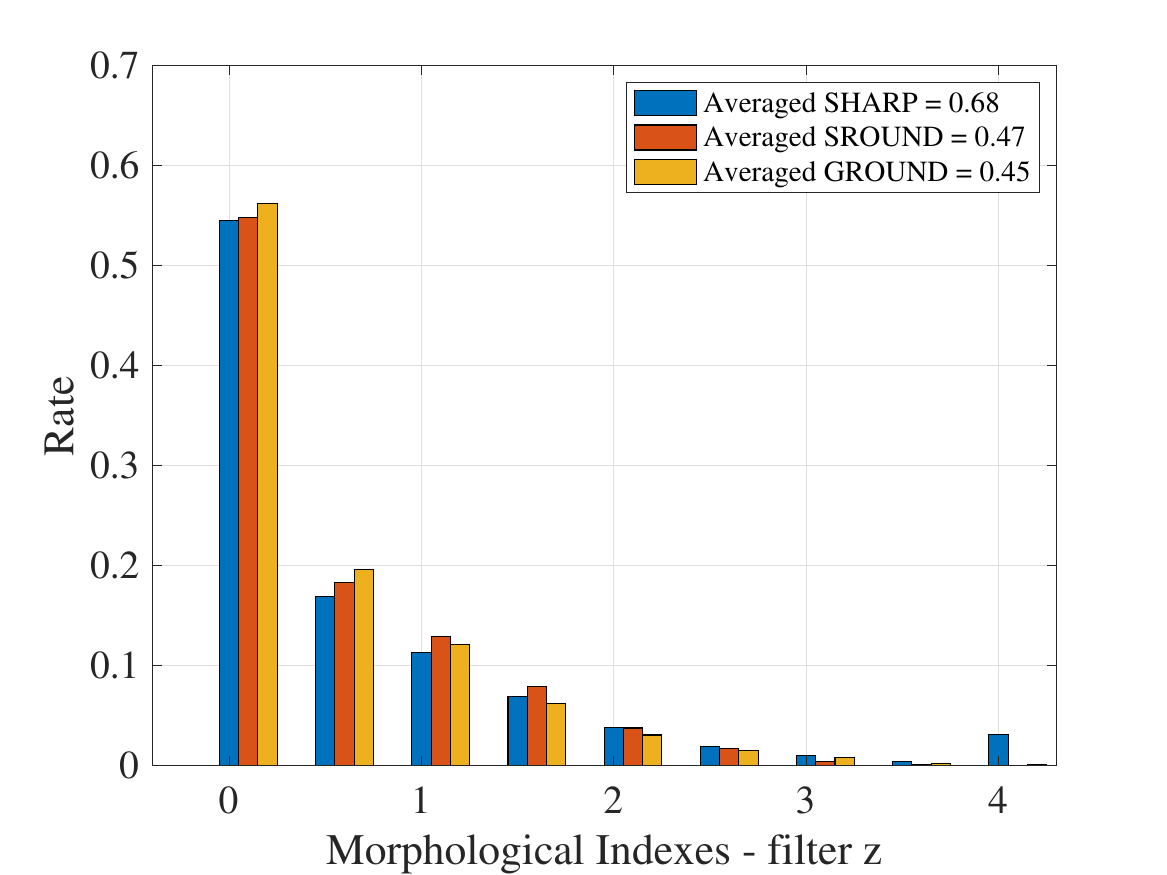}
	\small(H)
\end{tabular}
	\caption
	{Histogram of the morphological indexes from images from the SDSS. The left panels represent the morphological indexes of the point sources, and the right panels represent  the extended source. From top to bottom, followed by g, r, i, z filters. The average morphological indexes are labeled in the upper right corner of each figure.}
	\label{morph}%
\end{figure}

After matching QCC with SDSS DR16, we obtain 663,087 common sources, among which, 39,785 sources are identified as extended sources by SDSS. We randomly selected 18,400 extended sources and 18,400 point-like sources. See Fig. \ref{morph}, we plot the percentage histograms of the SHARP, SROUND, and GROUND\footnote{SHARP represents the ratio of, the difference between the height of the center pixel of the PSF and the mean of the surrounding non-bad pixels, to the height of the best fit Gaussian function at that point. SROUND computes the ratio of a measure of the bilateral symmetry of the object to a measure of the four-fold symmetry of the object. GROUND measures the ratio of, the difference in the height of the best fitting Gaussian function in x minus the best fitting Gaussian function in y, over the average of the best fitting Gaussian functions in x and y (\url{https://photutils.readthedocs.io/en/stable/api/photutils.detection.DAOStarFinder.html}).} morphological indexes of these sources. The median values of the morphological indexes of the extended sources in all four bands is slightly larger than that of the point-like sources as expected. In addition, both point-like and extended sources in all bands have small morphological indexes, only about 8\% of the sources have a morphological index greater than 2, which means most of them are stellar-like sources.

\subsection{Parallax, Magnitude and Proper motion}
\label{sec:Parallax and Magnitude}
As mentioned above, there are 1,186,690 sources in QCC that are also present in the $Gaia$ EDR3 AGN catalog, whose astrometric properties can be found in \citet{liao2021probing1}. Therefore, the astrometric properties of the remaining 316,683 newly identified quasar candidates are worth investigating. Based on this goal, we divided the QCC catalog into two subsamples: the QCC-A subset consisting of 1,186,690 quasar candidates already identified by EDR3, and the QCC-B subset consisting of the newly identified 316,683 quasar candidates. Among QCC-B subset, there are 113,186 (36\%) five-parameter sources, and 203,497 (64\%) six-parameter sources. After cross-matching  QCC-B with other AGN catalogs, we found that 106,928 sources have been identified as quasars (or quasar candidates), and the remaining 209,755 sources are newly identified.

\begin{figure}[htbp]
	\centering
	\includegraphics[width=0.45\textwidth]{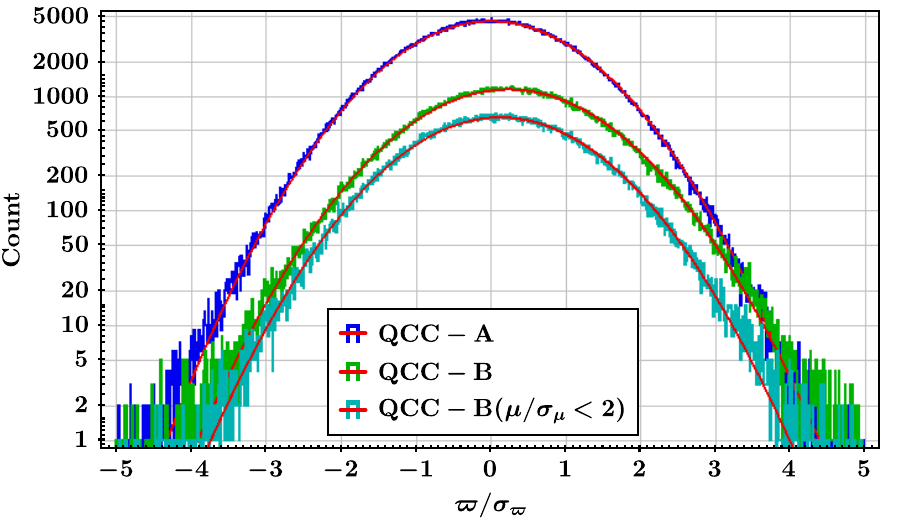}
	\caption{The normalized parallax distribution for the sources in QCC. The y-axis has been logarithmized, each bin of x-axis is 0.01 mas.  The red lines represent the best-fit Gaussian distributions.}
	\label{parallax}
\end{figure}

\begin{figure}[htbp]
	\centering
	\includegraphics[width=0.45\textwidth]{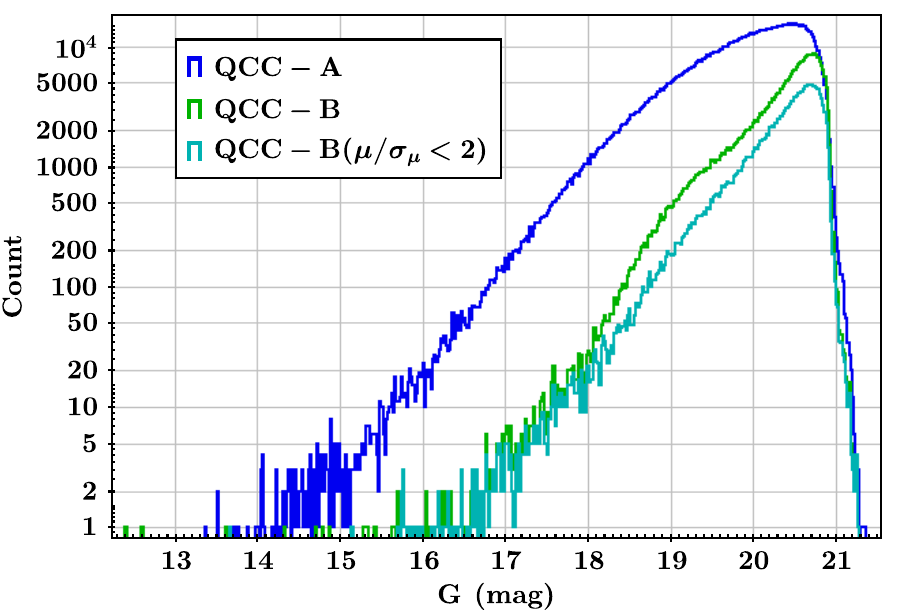}
	\caption{G magnitude distribution for the sources in QCC. The y-axis has been logarithmized, each bin of x-axis is 0.02 mag.}
	\label{Gmag}
\end{figure}

\begin{figure}[h]
	\centering
	\begin{tabular}{c}
		\includegraphics[width=0.6\textwidth]{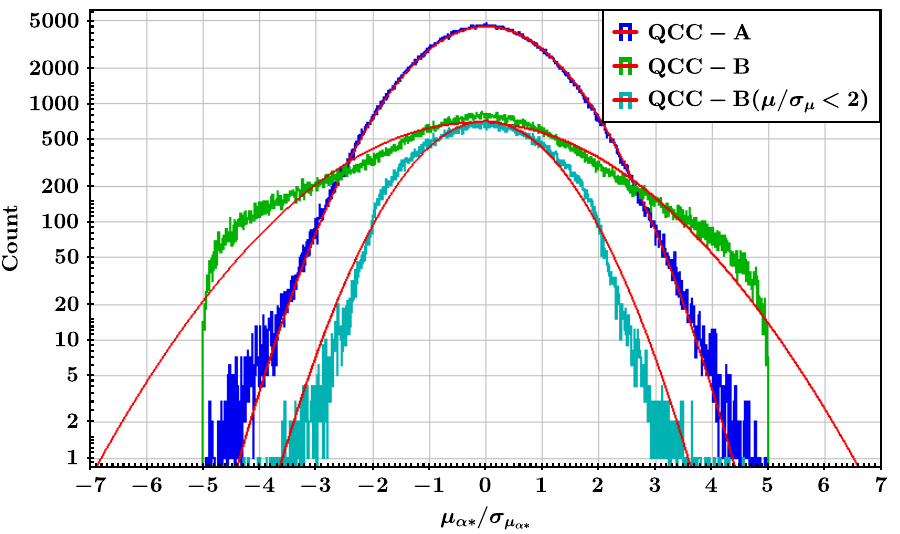}
		\small(A)
	\end{tabular}
	\begin{tabular}{c}
		\includegraphics[width=0.6\textwidth]{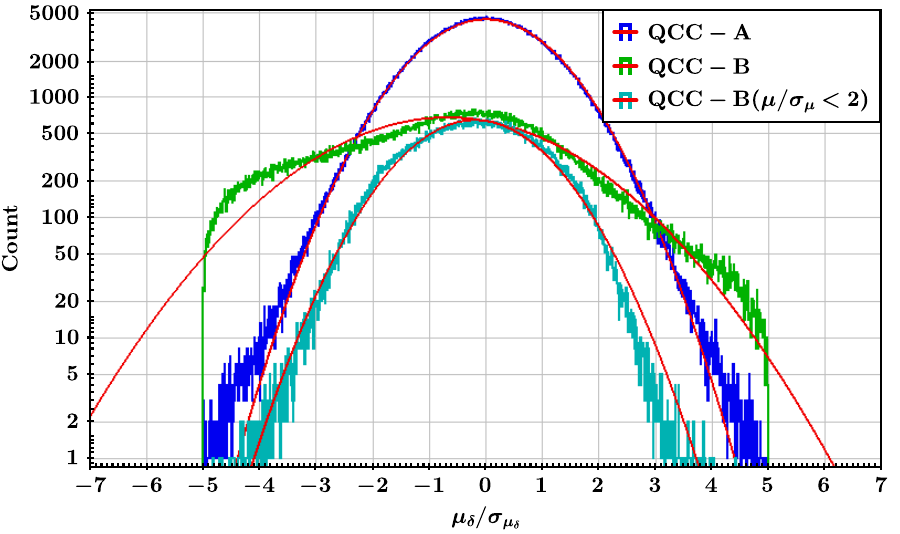}
		\small(B)
	\end{tabular}
	\caption
	{The normalized proper motions distributions for the sources in QCC. The red lines represent the best-fit Gaussian distributions.}
	\label{pm}%
\end{figure}

\begin{figure}[htbp]
	\centering
	\includegraphics[width=0.55\textwidth]{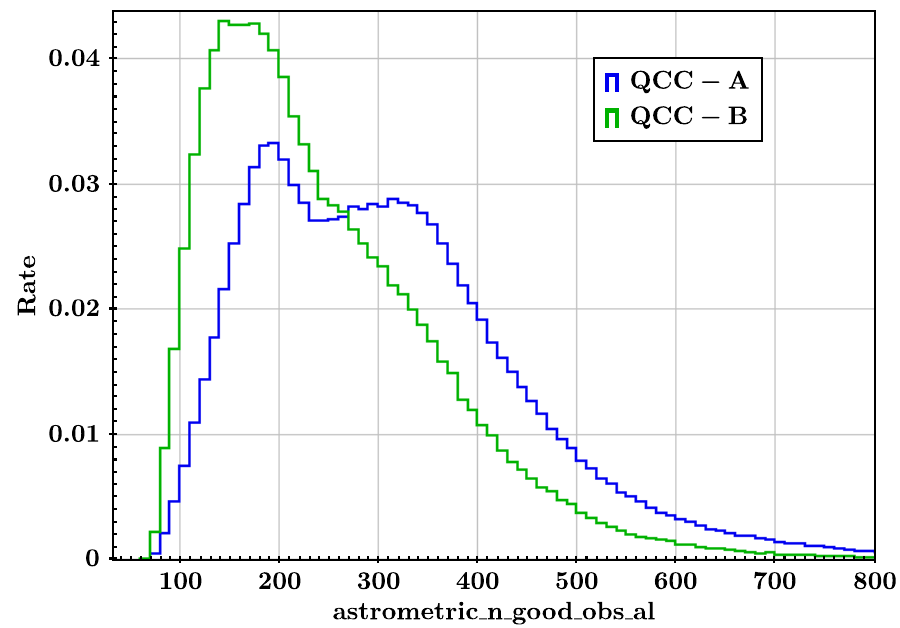}
	\caption{Number of good AL observations distribution for the sources in QCC. The y-axis represents the proportion of the number in each bin to the total sample, each bin of x-axis is 10.}
	\label{ngood}
\end{figure}

\begin{figure}[h]
	\centering
	\begin{tabular}{c}
		\includegraphics[width=0.7\textwidth]{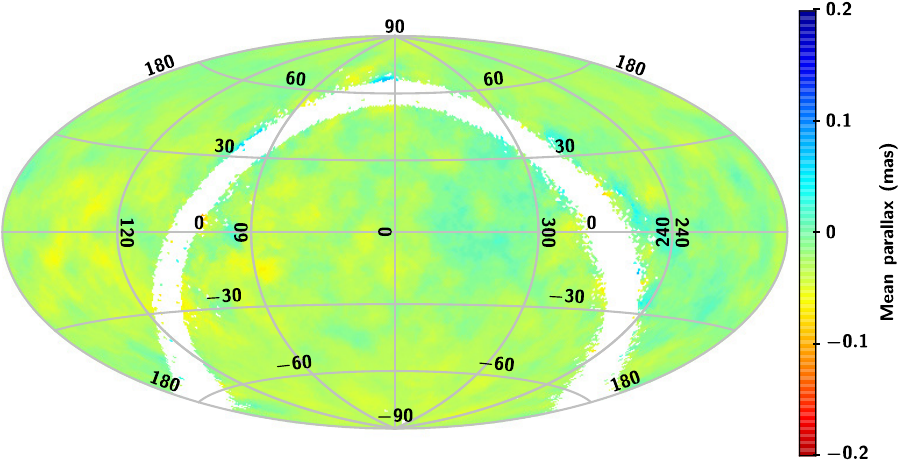}
		\small(A)
	\end{tabular}

	\begin{tabular}{c}
		\includegraphics[width=0.7\textwidth]{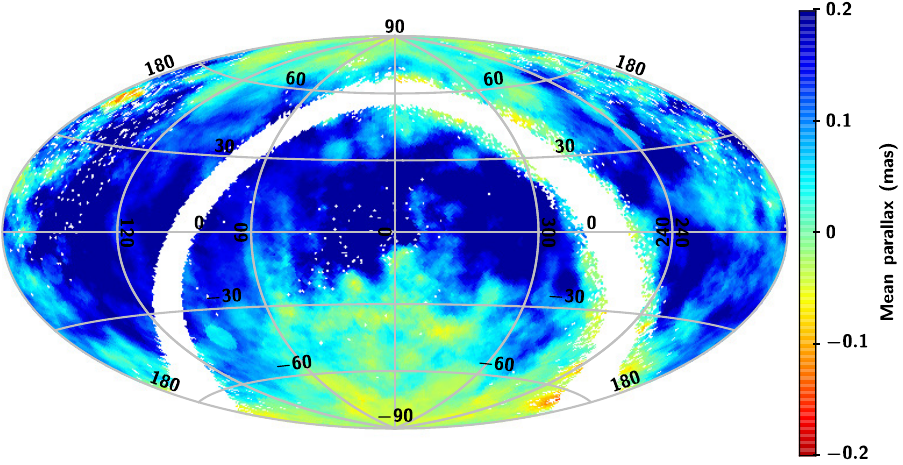}
		\small(B)
	\end{tabular}
	\caption
	{The generalized moving mean parallaxes of sources in QCC-A (A) and QCC-B (B). The map uses the Hammer Aitoff projection in Ecliptic coordinates.}
	\label{meanparallax}%
\end{figure}

\begin{figure}[htbp]
	\centering
	\includegraphics[width=0.65\textwidth]{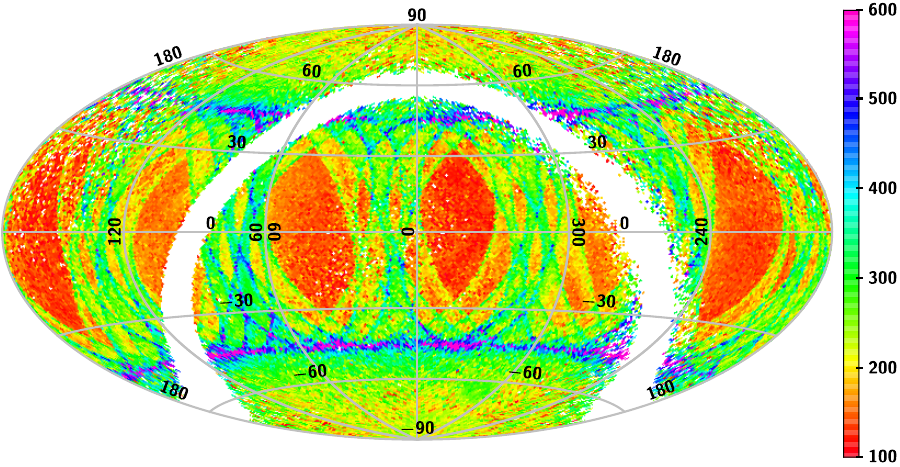}
	\caption{The good AL observations of sources QCC-B. The map uses the Hammer Aitoff projection in Ecliptic coordinates.}
	\label{ngoodqccn}
\end{figure}

The parallax, G magnitude and proper motion distribution can be found in Fig. \ref{parallax}, Fig. \ref{Gmag} and Fig. \ref{pm}, respectively. The sources in QCC-B populate the dimmer end. The median of magnitude of the QCC-B sample is 20.49 mag, while for the QCC-A sample, the median of magnitude is 20.00 mag. The average proper motion and parallax are shown in Table \ref{parameter}. The mean parallax and $\mu_{\delta}$ of QCC-B are significantly different from other quasar candidates, and the standard deviations are all very large. One possible reason is that these sources are fainter and less observed, see Fig. \ref{ngood}. As we proposed in \citep{liao2021probing1}, the number of good CCD observations along-scan greatly affects the astrometric solution of quasar candidates and cause a bias in the proper motion, especially for the six-parameter sources. Fig. \ref{meanparallax} shows the generalized moving mean\footnote{The generalized moving mean used the neighboring points on the celestial sphere to smooth each point by using a generalized weighting function  \citep{bucciarelli1993two}. To be compared with the smoothed maps and median parallax plot in $Gaia$ EDR3  \citep{lindegren2021gaia,fabricius2021gaia}, the generalized moving mean also calculated each source in a $5^\circ$ radius region and with more than 50 objects.} (GMM hereafter) parallaxes of sources in QCC, the parallax distribution of sources in QCC-A is relatively uniform, but for QCC-B, sources locate within $\pm 30^{\circ}$ of the ecliptic plane show significant parallax bias. We find most of these sources have less than 200 good AL observations, see Fig. \ref{ngoodqccn}.

\begin{figure}[h]
	\centering
	\begin{tabular}{c}
		\includegraphics[width=0.4\textwidth]{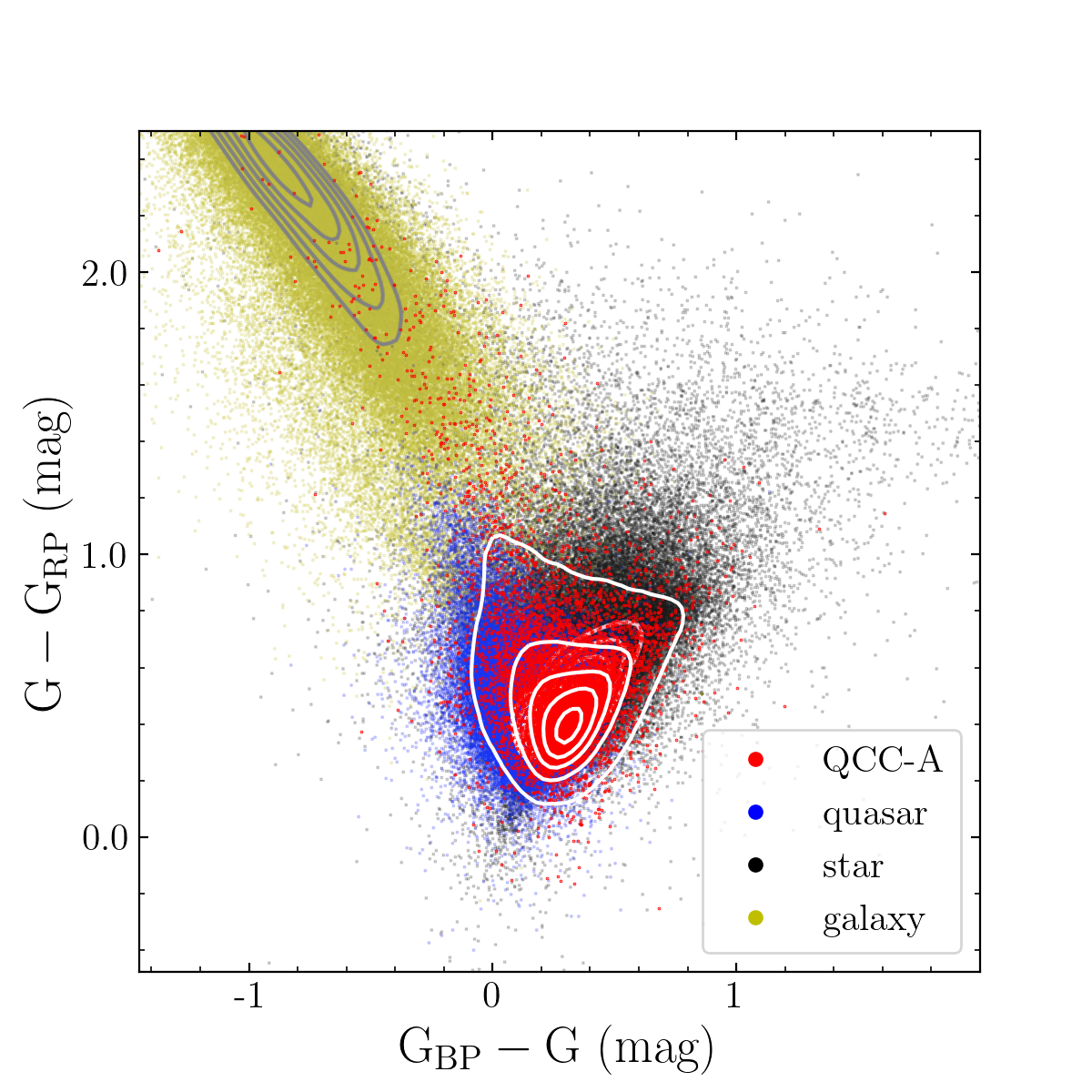}
		\small(A)
	\end{tabular}
	\begin{tabular}{c}
		\includegraphics[width=0.4\textwidth]{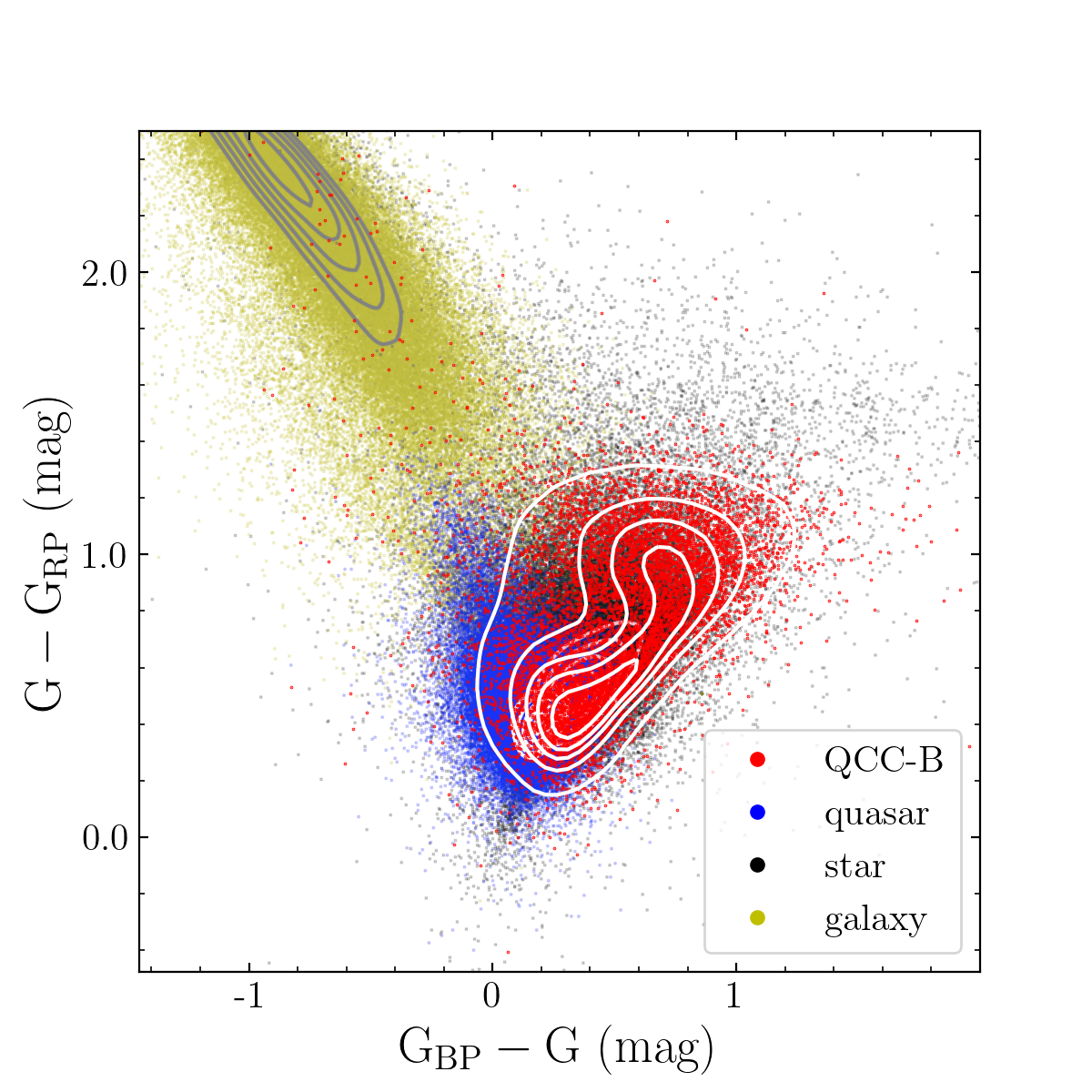}
		\small(B)
	\end{tabular}

	\begin{tabular}{c}
	\includegraphics[width=0.4\textwidth]{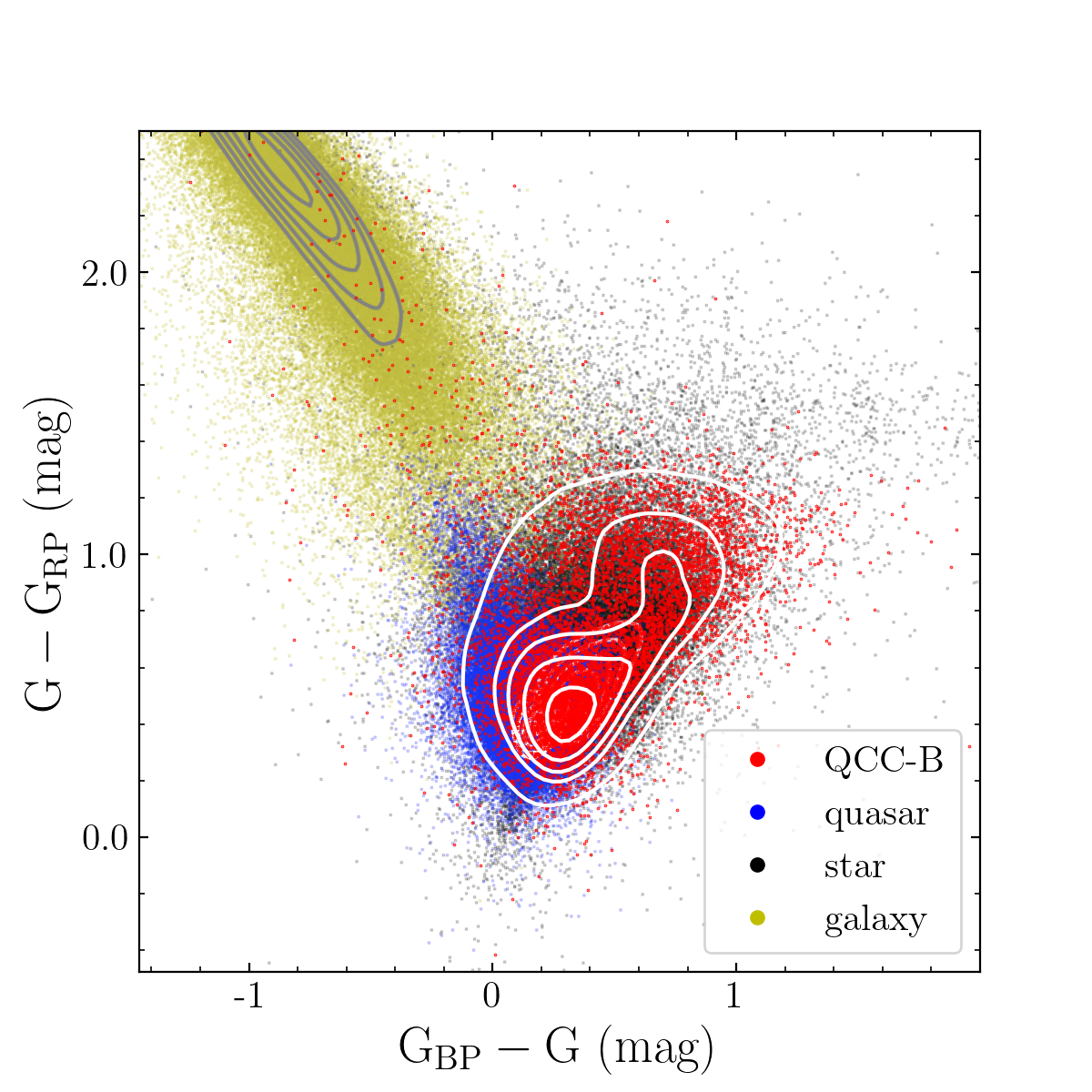}
	\small(C)
\end{tabular}
\begin{tabular}{c}
	\includegraphics[width=0.4\textwidth]{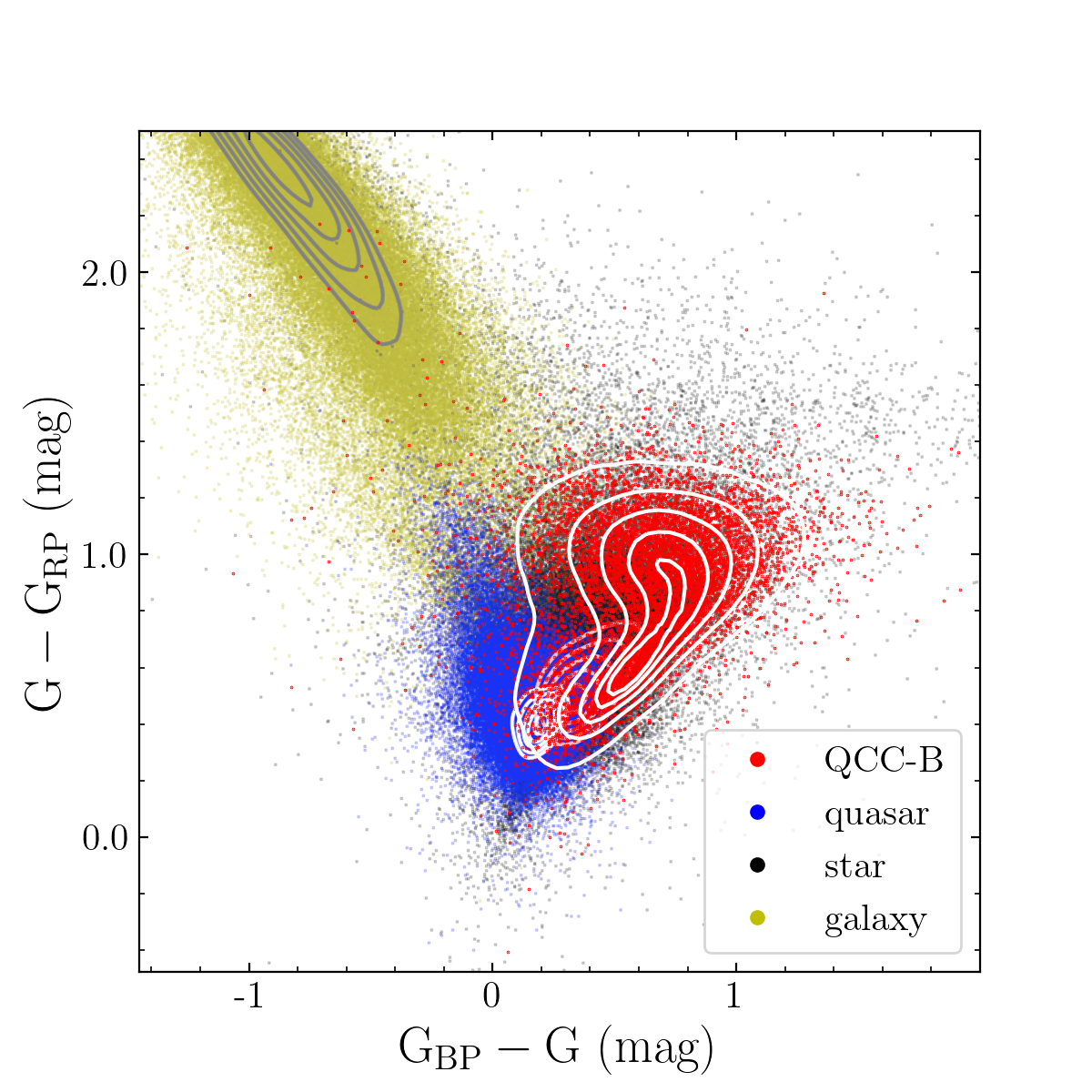}
	\small(D)
\end{tabular}
	\caption
	{The colour–colour diagram for the sources in QCC-A (A) and QCC-B (B). Figure (C) and (D) are the color distribution of sources in QCC-B ($\mu/\sigma_{\mu} \leq 2$) and QCC-B ($\mu/\sigma_{\mu} \textgreater 2$), respectively. The red dots represent 10,000 randomly selected sources in each sample, and the contours in each figure show the variation in source density of the whole sample on a linear scale.}
	\label{color}%
\end{figure}

Another cause of the parallax and proper motion bias might be stellar contamination. \citet{bailer2019quasar} constructed a supervised classifier based on Gaussian Mixture Models to probabilistically classify extragalactic objects in $Gaia$ DR2. See Fig. \ref{color}, we plot the colour–colour diagram for our selection to compare with the training set coloured according to the true classes used in \citet{bailer2019quasar}. We find that for the QCC-B subset, some sources locate at the quasar-star overlap area, which indicates the presence of star contamination. A rough estimate shows that the purity of QCC-B is about 42.2\%-58.5\% \footnote{Two ways are implemented for the purity testing. A): we cross-matched QCC-B with SDSS DR16 and found that the purity of QCC-B was 58.5\%. B):  \citet{collaboration2022gaia} provided a low purity (50\%-70\%) QSO candidate table with 6.6 million sources (QCT hereafter) and a QSO sub-sample with 95\% purity (QSS hereafter). For the 316,683 sources in QCC-B, 54,343 sources are found in QSS, 38,266 sources are found in QCT but not in QSS. Then we matched the remaining 224,074 sources with SDSS DR16, and the purity of these sources is about 28\% according to the spectroscopic classifications of 79,154 common sources. Based on these results, the purity of QCC-B is estimated to about 42.2\%-44.6\%. Using the same methods,  the purity of the overall QCC is about 87.8\%-91.3\%, which is consistent with our conclusion in Section \ref{reliab}.}. 

According to above investigation, there is obvious stellar contamination in QCC-B, which will lead to confusion when using QCC. So it is necessary to further reduce the stellar contamination and select purer sample. \citet{heintz2018unidentified} came up with an idea that $S/N_{\mu} = \mu/\sigma_{\mu} \textless 2$ is a more strict and effective criterion for identifying quasars. We investigated the $\mu/\sigma_{\mu}$ distribution of the quasars in $Gaia$ EDR3 AGN catalog, 90\% of these quasars have a $\mu/\sigma_{\mu}$ less than 2. After applying this criteria to QCC-B, we find the mean parallax and proper motion bias is significantly reduced, see Table \ref{parameter}. Moreover, the standard deviations of the best-fit Gaussian distributions in Fig. \ref{parallax} are 1.055, 1.106 and 1.074 for the QCC-A, QCC-B and QCC-B ($\mu/\sigma_{\mu} \textless 2$) subset, respectively. In Fig. \ref{pm}, the corresponding standard deviations of the best-fit Gaussian distributions are 1.062, 1.840 and 0.990 for $\mu_{\alpha*}$, and 1.072, 1.870 and 1.087 for $\mu_{\delta}$, respectively. Stricter criteria of proper motions significantly reduce the standard deviations of the best-fit Gaussian distributions of parallax and proper motion, and the distribution of the sources with $\mu/\sigma_{\mu} \textless 2$ is closer to the quasar region in Fig. \ref{color}, which indicates a more reliable quasar candidates. To further reduce stellar contamination in the sample, we select 99,673 sources with $G_{BP}-G \leq 0.8$ and   $G-G_{RP} \leq 0.8$ in the QCC-B ($\mu/\sigma_{\mu} \textless 2$) subset. And their average parallax, $\mu_{\alpha*}$ and $\mu_{\delta}$ are 1.6$\pm$48.2 $\mu as$, 0.2$\pm$38.7 $\mu as/yr$ and -15.3$\pm$41.0 $\mu as/yr$ respectively, which indicates smaller bias and residuals in these astrometric parameters. We have also marked these sources as reliable quasar candidates (RQC) in our catalog, and the purity of RQC we estimated was about 73.1\%-85.2\% \footnote{Referring to the approach to estimate the purity of QCC-B in this section, the purity of RQC estimated by method A and B is 85.2\% and 73.1\%-78.9\%, respectively.}.


\begin{table*}[!t]
	\centering
	\caption{Comparison of parameters between $Gaia$ EDR3 AGN and QCC, all averages are derived from data after GMM.}
	\begin{tabular}{ccccc}
		\hline
		Parameters &$Gaia$ EDR3 AGN & QCC-A & QCC-B & QCC-B $(\mu/\sigma_{\mu} \textless 2)$ \\ \hline
		Number &1614173& 1186690&316683&175168 \\ 
		Median of Gmag (mag) &20.06&20.00&20.49&20.49 \\ 
		Weighted average of parallax ($\mu$as)&-21.8$\pm$9.6&-22.0$\pm$10.1&90.4$\pm$87.1&33.2$\pm$71.0\\ 
		Weighted average of $\mu_{\alpha_*}$ ($\mu$as/yr)&-0.3$\pm$11.3&-0.6$\pm$11.5&-2.5$\pm$247.8& 0.5$\pm$67.0\\ 
		Weighted average of $\mu_{\delta}$ ($\mu$as/yr)&-1.5$\pm$10.4&-1.1$\pm$10.4&-224.2$\pm$254.1&-51.4$\pm$90.2\\ \hline
	\end{tabular}\label{parameter}
\end{table*}

\section{Discussion}
\label{discussion}
\subsection{Quasar candidates in the Galactic plane}
\label{gpq}
As seen in Eq \ref{criteria}, to lower the possibility of stellar contamination, the quasar-like objects identified in $Gaia$ EDR3 and QCC have ruled out the objects within the  Galactic plane ($\lvert \sin b \rvert \leq 0.1$). The coolest brown dwarfs and the most heavily dust-reddened stars will exhibit similar WISE colors as quasars near the Galactic plane \citep{stern2012mid}. \citet{kirkpatrick2011first} showed that stars of spectral class later than T1 dwarfs have W1-W2 $\geq$ 0.8 mag, which means that the color selection criterion from the WISE data is not working effectively near the Galactic plane.  The most reliable way to identify a quasar near the Galactic is the spectrum method, such as LAMOST Spectroscopic Survey of the Galactic Anti-center (LSSGAC)  \citep{liu2013lss}. With the spectral data, \citet{huo2017quasars} presented a sample of 151 quasars discovered in an area near the Galactic Anti-Center. Machine learning is another important method. \citet{fu2021finding} synthesized quasars and galaxies behind the Galactic plane and applied the XGBoost algorithm to Pan-STARRS1 (PS1) \citep{flewelling2020pan} and AllWISE photometry for quasar classification, in which they obtained the Quasars behind the Galactic Plane (GPQ) candidate catalog with 160,946 sources located at $\lvert b \rvert $ $\leq$ $20^{\circ}$. Since the quasars are very important, under the situation of lacking spectrum data, we intend to check the reliability of these quasar candidates by analyzing their astrometric solutions.

Cross matching near the Galactic Equator is heavily affected by the confusion sources, which may lead to problematic cross-matching \citep{refId0}. Therefore, we carefully matched with the PS1 source\_id provided by GPQ and the $gaiadr3.panstarrs1\_best\_neighbour$ provided by $Gaia$ Archive. Note that although this will improve the accuracy of the match, the problematic matching  may still exist. We obtained 133,798 common sources of GPQ and $Gaia$ EDR3. Among them, there are 76,225 five-parameter sources and 37,684 six-parameter sources  in the magnitude range from 6 to 21 Gmag.  For comparison,  we have selected the quasar candidates near the Galactic plane ($\lvert b \rvert \textless10^{\circ}$) in FRS,  all candidates from EDR3\_AGN catalog and spectroscopic confirmed quasars from LAMOST DR7 (LAMOST DR7Q hereafter) (\url{http://dr7.lamost.org/v2.0/}). See Table \ref{tab:corrparallax} and Table \ref{tab:4pm} for the mean parallax and proper motion of each quasar sample. Except for GPQ, the mean parallaxes of five-parameter sources for each sample are consistent with each. Additionally, for the six-parameter sources,  the mean parallaxes of GPQ and LAMOST DR7Q are obviously different from EDR3\_AGN catalog. And for the mean proper motions of these samples, GPQ is evidently different from other catalogs in both 5-parameter and 6-parameter sources. 
Assuming that quasars in different Galactic latitudes have similar astrometric system errors, the significant positive mean parallax and negative mean proper motion of GPQ six-parameter sources might be caused by stellar contamination. We will discuss the systematic errors in different sky regions in detail in Appendix \ref{system error}.

The $Gaia$ team provided a parallax bias correction model,  which proposes that the parallax bias is at least related to the g-band magnitude G, ecliptic latitude $\beta$ and photometric parameter $\nu_{eff}$. For the faint sources, this model is derived from the quasar candidates of EDR3 AGN  \citep{lindegren2021gaia1}.  To investigate the effectiveness of this model in the Galactic plane,  we apply it to the GPQ and LAMOST DR7Q quasar samples. Again, we use FRS and EDR3\_AGN catalog as comparisons. See Table \ref{tab:corrparallax}, for the five-parameter sources, the parallax biases of the FRS and LAMOST DR7Q samples are corrected to about 5 $\mu as$, while 26.5 $\mu as$ and 0.5 $\mu as$ for the GPQ and EDR3 AGN catalog, respectively. For the six-parameter sources, the parallax biases of GPQ and LAMOST DR7Q samples deviate significantly from zero, suggesting the correction model is not working effectively.  These results indicate that (i) The crowded sky near the Galactic Equator may have caused significant GPQ matching errors; (ii) There might be a stellar contamination with the GPQ sample,  especially in the six-parameter sources; (iii) Compared to high Galactic latitude regions, the photometric data obtained from the Galactic plane follow different probability distribution. Therefore, the parallax bias correction model provided by the $Gaia$ team is not working effectively in the Galactic plane.

\begin{table*}[!t]
	\centering
	\caption{Parallaxes of GPQs, $Gaia$-FRS, LAMOST DR7Q and $Gaia$ EDR3 AGN before and after correction.}
	\begin{tabular}{ccccccc}
		\hline
		&       &       & \multicolumn{2}{c}{ Parallax ($\mu$as)} & \multicolumn{2}{c}{ Parallax after correction ($\mu$as)} \\ \hline
		&       & Number & weighted average & median & weighted average & median \\ \hline
		\multirow{2}[0]{*}{GPQs} & 5-parameter & 76225  &5.6 & -6.7 & 26.5   & 10.4 \\
		& 6-parameter & 37684  & 25.0  & 13.7   & 43.7  & 32.9 \\ \hline
		$Gaia$-FRS ($\lvert b \rvert $ \textless $10^{\circ}$) & 5-parameter & 5844  & -19.4 & -15.0 & 5.4   & 5.7 \\ \hline
		\multirow{2}[0]{*}{LAMOST DR7Q} & 5-parameter & 38016  & -16.2 & -22.5 & 7.4   & -3.5 \\
		& 6-parameter & 4379  & -35.6 & -45.5 & -19.4 & -32.2 \\ \hline
		\multirow{2}[0]{*}{EDR3 AGN} & 5-parameter & 1215942  & -21.2 & -17.8 & 0.5   & 0.2 \\
		& 6-parameter & 398231  & -27.5 & -28.5 & -8.9  & -9.1 \\ \hline
	\end{tabular}%
	\label{tab:corrparallax}%
\end{table*}%

\begin{table*}[!t]
	\centering
	\caption{Proper motions of GPQs, $Gaia$-FRS, LAMOST DR7Q and $Gaia$ EDR3 AGN.}
	\begin{tabular}{ccccccc}
		\hline
		&       &       & \multicolumn{2}{c}{ $\mu_{\alpha*}$ ($\mu$as/yr)} & \multicolumn{2}{c}{$\mu_{\delta}$ ($\mu$as/yr)} \\ \hline
		&       & Number & weighted average & median & weighted average & median \\ \hline
		\multirow{2}[0]{*}{GPQs} & 5-parameter & 76225  & -10.1 & -15.4 & -80.1  & -8.4 \\
		& 6-parameter & 37684  & -222.0 & -61.5 & -382.8& -70.7 \\ \hline
		$Gaia$-FRS ($\lvert b \rvert $ \textless $10^{\circ}$) & 5-parameter & 5844  & 0.1   & 3.7   & 0.3   & 3.7 \\ \hline
		\multirow{2}[0]{*}{LAMOST DR7Q} & 5-parameter & 38016  & -3.1  & -2.0  & 0.4   & -1.9 \\
		& 6-parameter & 4379  & 2.4   & 6.4   & 4.3   & 3.0 \\ \hline
		\multirow{2}[0]{*}{EDR3 AGN} & 5-parameter& 1215942  & -0.3  & -0.1  & -1.3  & -1.3 \\
		& 6-parameter& 398231  & -0.6  & -2.2  & -4.4  & -4.0 \\ \hline
	\end{tabular}%
	\label{tab:4pm}%
\end{table*}%

\subsection{Quasar candidates with abnormal astrometric behavior}

There are lots of spectroscopically identified quasars contained in the catalogs listed in Table \ref{reliability} and Table \ref{tab:completeness}. However, the astrometric parameters of some quasars are significantly abnormal. For example, some quasars have large bias in proper motion and parallax, or have obvious position difference between $Gaia$ DR2 and EDR3. This indicates that $Gaia$'s high-precision observations could detect the jitter of these quasars, or as mentioned in \citet{shen2021hidden} and \citet{chen2022varstrometry}, they might be special quasars, such as quasar pairs and lensed quasars, which are important in exploring the evolution of galaxies and finding double black holes. On the other hand, these quasars are not suitable for establishing the celestial reference frame. Therefore, it is necessary to exclude them from the quasar candidates used for the celestial reference frame.

\begin{figure}[h]
	\centering
	\begin{tabular}{c}
		\includegraphics[width=0.4\textwidth]{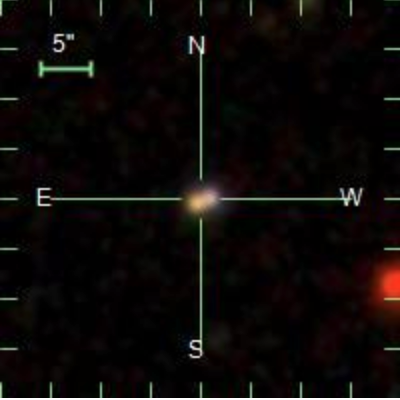}
		\small(A)
	\end{tabular}
	\begin{tabular}{c}
		\includegraphics[width=0.4\textwidth]{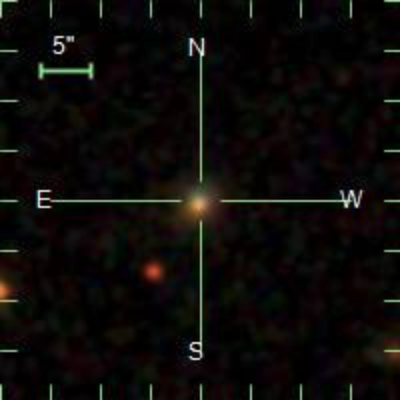}
		\small(B)
	\end{tabular}

	\begin{tabular}{c}
		\includegraphics[width=0.4\textwidth]{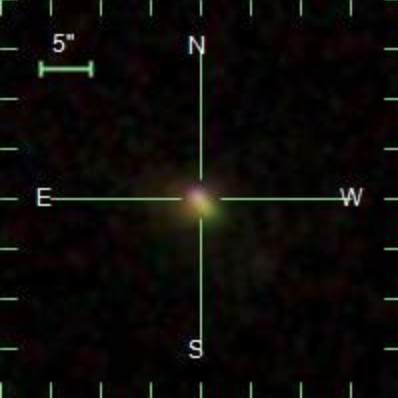}
		\small(C)
	\end{tabular}	
	\begin{tabular}{c}
		\includegraphics[width=0.4\textwidth]{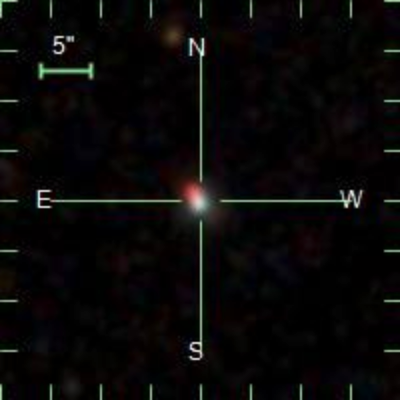}
		\small(D)
	\end{tabular}
	\caption
	{Four SDSS DR16 quasars images with abnormal astrometric behavior.}
	\label{badqso}%
\end{figure}

We made a preliminary attempt to select such abnormal astrometric quasars. We have selected SDSS quasars with more than one corresponding source in $Gaia$ EDR3 within $1^{\prime \prime}$. These quasars may be affected by nearby sources, therefore, they have large positional errors. For the sources with a single $Gaia$ matched within a $1^{\prime \prime}$ radius of the SDSS position, inspired by \citet{lindegren2021gaia}, we have selected some abnormal astrometric quasar candidates based on several astrometric parameters. These parameters with large values indicate a bad fit of each source or probably are unsolved double stars. We checked the SDSS images of these spectroscopically confirmed quasars with such astrometric behavior and found that most of them are extended-source obviously. Fig. \ref{badqso} shows four sources of such quasars: (A) and (B) are SDSS quasars with two $Gaia$ matches within $1^{\prime \prime}$ radius, while (C) and (D) are two arbitrary quasars among sources selected based on the above parameters. Among them, Fig. \ref{badqso} (D) is an image of J013958.43+321631.6, whose corresponding values are 10.673754 mas, 5.370112 for $astrometric\_excess\_noise$ and $astrometric\_excess\_noise\_sig$, respectively. Although most of them have near-zero parallaxes and proper motions, their positions are not reliable due to the large astrometric jitter observed by $Gaia$ or the presence of other sources very close to them ($\textless$ 1 arcsec). Therefore, we should remove these sources when establishing the celestial reference frame. More details could be found in \citep{10.3389/fspas.2022.822768}. Based on this study, we have found 284 abnormal astrometric quasars in QCC and flagged them in our catalog.

\subsection{Limitation of this work}

Selecting quasar candidates with astrometric and mid infrared data has been proved to be effective and reliable. Based on this, we have selected millions of quasar candidates with Gaia and AllWISE data. These quasar candidates with high completeness and purity will be important in the process of realizing the celestial reference frame. 

However, there are still some limitations in our approach. First of all, as we emphasized in Section \ref{sec:Parallax and Magnitude}, Gaia EDR3 use only 34 months observations, which greatly affects the astrometric solutions. These results will affect the purity of the quasar sample selected by astrometric methods. With longer observations in the future, this situation will be improved. Secondly, the number of quasars identified by this method heavily depend on the number of objects with infrared data provided by AllWISE, which means our quasar candidate catalog ruled out many sources that not been observed by AllWISE. As discussed in Section \ref{reliab}, the completeness and purity will decrease with fewer WISE observation. Therefore, our quasar sample is less complete and pure than the quasar candidates provided by Gaia \citep{refId0}. Thirdly, due to the heavy dust in the crowded areas such as the Galactic plane and LMC/SMC, the WISE color loses  its effectiveness to distinguish brown dwarfs and the dust-reddened stars from quasars. In such cases, the selection results with our method should be used with caution. Further efforts are required to identify quasars in these crowded areas. Implementing spectroscopic surveys to these areas is the most reliable way. And using machine learning methods with explicit account for the Galactic extinction and reddening to provide external catalogues specially for the Galactic plane is also a feasible approach.

As claimed by \citet{hog2014absolute}, astrometric detection of quasars, i.e. to identify quasars only from the characteristics of zero proper motion and parallax, which is unbiased by any assumptions on spectra, might lead to discovery of a new kind of extragalactic point sources \citep{heintz2015study}. This issue can be verified with more $Gaia$ observations and more accurate astrometry data in the future.


\section{Summary and Conclusions}
\label{sec:Conclusions}

Quasars are one type of active galactic nuclei. Because of their bright centers and point-like appearances, quasars are the perfect objects to establish the celestial reference frame. The 3rd realization of the International Celestial Reference Frame (ICRF3) is established by 4588 sources at radio wavelengths, there are about 22\% of sources show great offset in optical and radio positions   \citep{charlot2020third}. With the high-precision astrometric parameters for more than 1.8 billion sources of $Gaia$ Early Data Release 3 (EDR3), lots of quasar candidates can be identified. To establish a non-rotating celestial reference frame in the optical band, we need a reliable catalog with a large number of quasars. In this paper, we  used the astrometry data of $Gaia$ EDR3 and color data of AllWISE to identify quasar candidates and made a comprehensive evaluation of them.

A quasar candidate catalog (QCC) of 1,503,373 sources (about 90\% purity) is obtained by astrometric and mid-infrared methods in $Gaia$ EDR3, which has 1,186,690 (78.9\%) candidates in common with $Gaia$ EDR3's AGN catalog. The purity of 316,683 newly identified quasar candidates is about 42.2\%-58.5\%. Compared with LQAC5, the completeness of our catalog is around 80\%, and we randomly select 4171 common sources of our catalog and SDSS DR16, according to the spectrums of SDSS, about 99.7\% of which are quasars, 94.0\% are point-like sources. Compared to the previous similar research \citep{guo2018identifying}, we have selected more quasar candidates (1,503,373 versus 662,753) with higher purity (90\% versus 77\%). Star contamination is present in the newly identified subset and the purity of this subset improved significantly after we used more stringent astrometric and color conditions. In addition, we found that the purity of quasar candidates selected by mid-infrared and astrometric data decreases around the LMC/SMC, area near the Galactic Equator and at the fainter magnitude.

We find that the parallax correction model of $Gaia$ EDR3 cannot be directly applied to sources near the Galactic plane, especially to the 6-parameter sources. We also select the quasars with abnormal astrometric behavior, which are not suitable for establishing the celestial reference frame and should be excluded from the quasar candidates for such purpose. We can foresee that with the future release of $Gaia$ data, the identification of quasars using astrometric methods will have increasing reliablity. 

Although $Gaia$ have provided more than six millions quasar candidates in $Gaia$ DR3 \citep{collaboration2022gaia}, the reliability of the quasar candidate list needs to be tested by the quasar catalog obtained by other methods. The quasar candidate catalog obtained by astrometric and mid-infrared methods will play an essential role to verify the future release of $Gaia$ data.

\begin{acknowledgements}
	We thank the anonymous reviewers for their valuable suggestions. We used data from AllWISE to achieve this work: AllWISE makes use of data from WISE, which is a joint project of the University of California, Los Angeles, and the Jet Propulsion Laboratory/California Institute of Technology, and NEOWISE, which is a project of the Jet Propulsion Laboratory/California Institute of Technology, WISE and NEOWISE are funded by the National Aeronautics and Space Administration. And this work has made use of data from the European Space Agency (ESA) mission {\it $Gaia$} (\url{https://www.cosmos.esa.int/gaia}), processed by the {\it $Gaia$} Data Processing and Analysis Consortium (DPAC,
	\url{https://www.cosmos.esa.int/web/gaia/dpac/consortium}). Funding for the DPAC has been provided by national institutions, in particular the institutions participating in the {\it $Gaia$} Multilateral Agreement. We are also very grateful to the developers of the TOPCAT \citep{taylor2005topcat} software. This work has been supported by the Youth Innovation Promotion Association CAS, the grants from the Natural Science Foundation of Shanghai through grant 21ZR1474100, and National Natural Science Foundation of China (NSFC) through grants 12173069, and 11703065. We acknowledge the science research grants from the China Manned Space Project with NO.CMS-CSST-2021-A12 and NO.CMS-CSST-2021-B10. 
\end{acknowledgements}

\bibliographystyle{raa} 
\bibliography{hhbib}

\begin{appendix}

\section{The astrometric system error of different sky regions}
\label{system error}
\begin{figure}[h]
	\centering
	\begin{tabular}{c}
		\includegraphics[width=0.6\textwidth]{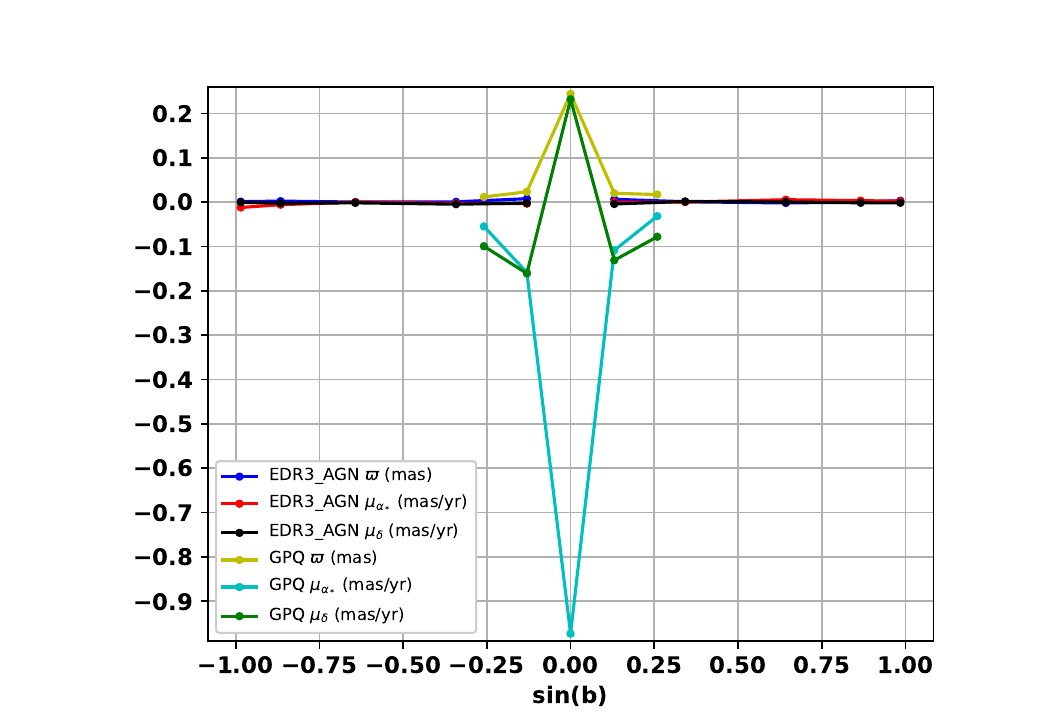}
		\small(A)
	\end{tabular}
	\begin{tabular}{c}
		\includegraphics[width=0.6\textwidth]{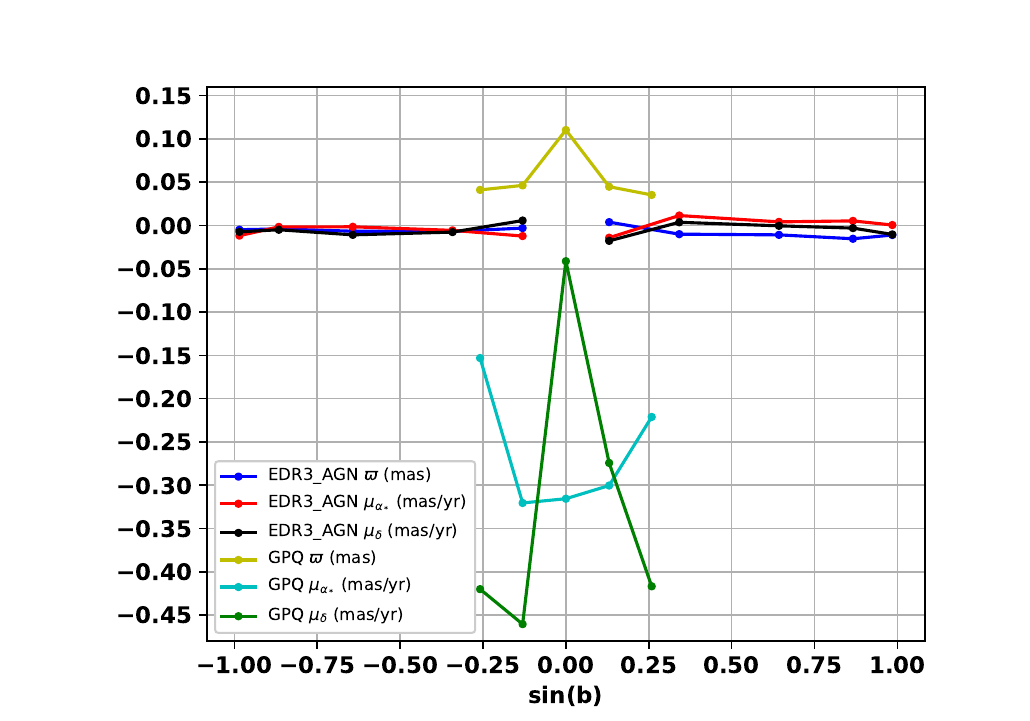}
		\small(B)
	\end{tabular}
	\caption
	{The astrometric system error of different Galactic latitude for 5-parameter sources (A) and 6-parameter sources (B).}
	\label{error_gpq}%
\end{figure}

In Section \ref{gpq}, we found possible stellar contamination in GPQ, especially the six-parameter sources in this catalog. These conclusions are based on the assumption that sources in different sky regions possess similar systematic errors.  The sources in GPQ are close to the Galactic Equator, so we investigated the variation of astrometric system error with Galactic latitude. We found that most of the sources in GPQ are located at the region of $\lvert b \rvert \textgreater5^{\circ}$, and only 3993 (3.51\%) sources have a Galactic Latitude less than $5^{\circ}$ (and larger than $-5^{\circ}$). This means that the GPQ and EDR3\_AGN catalogs have some overlapping sky regions, which is $20^{\circ}\leq \lvert b \rvert \leq5^{\circ}$ (overlap-region here after). Therefore, we calculated the mean proper motion and corrected parallax of sources at different Galactic latitudes. Fig. \ref{error_gpq} shows the distributions of astrometric system errors. Firstly, both 5-parameter and 6-parameter sources with $\lvert b \rvert \textless5^{\circ}$ in GPQ have a obvious systematic errors. The number of sources in these regions is very small (about 3,000), which makes the average value of system errors easy to be significantly affected by some extreme values, so these system errors might be unreliable. Secondly, at the overlap-region, the systematic errors of most sources in GPQ is larger than that of sources in EDR3\_AGN, especially for 6-parameter sources. Thirdly, for EDR3\_AGN, the systematic error of the sources has no obvious change at high Galactic latitude.

\end{appendix}

\label{lastpage}
\end{document}